\documentclass[reprint,amsmath,amssymb,aps]{revtex4-2}
\usepackage{graphicx}
\usepackage{amsmath,amssymb}
\usepackage{ascmac}
\usepackage{url}
\usepackage{layout}
\usepackage{color}
\usepackage{braket}
\usepackage{physics}
\usepackage{mathrsfs}
\usepackage{booktabs}
\usepackage{here}
\usepackage{listings}
\usepackage{dcolumn}
\usepackage{bm}

\def\lesssim{\ \raise.3ex\hbox{$<$}\kern-0.8em\lower.7ex\hbox{$\sim$}\ }
\def\gesim{\ \raise.3ex\hbox{$>$}\kern-0.8em\lower.7ex\hbox{$\sim$}\ }

\begin{document}
\title{Pairing properties of an odd-frequency superfluid Fermi gas}
\author{Shumpei Iwasaki}
\author{Taira Kawamura}
\author{Koki Manabe}
\author{Yoji Ohashi}
\affiliation{Department of Physics, Keio University, 3-14-1 Hiyoshi Kohoku-ku, Yokohama 223-8522, Japan}
\begin{abstract}
We theoretically investigate strong-coupling properties of an odd-frequency Fermi superfluid. This pairing state has the unique property that Cooper pairs are formed between fermions, not at the same time, but at different times. To see whether or not such unequal-time pairs still exhibit bosonic behavior, we examine the space-time structure of the odd-frequency Cooper-pair wavefunction at $T=0$, by employing the combined path-integral formalism with the BCS-Eagles-Leggett-type superfluid theory. In the strong-coupling regime, the odd-frequency pair wavefunction still has different space-time structure from that in the ordinary even-frequency $s$-wave superfluid state, their {\it magnitudes} are found to become close to each other, except for the equal-time pairing component. In this regime, we also evaluate the superfluid phase transition temperature $T_{\rm c}$ within the framework of the strong-coupling theory developed by Nozi\`eres and Schmitt-Rink. The calculated $T_{\rm c}$ in the strong-coupling regime of the odd-frequency system is found to be well described by the Bose-Einstein condensation of tightly bound Bose molecules. Our results indicate that, in spite of vanishing equal-time pairing, odd-frequency Cooper pairs still behave like bosons in the strong-coupling regime, as in the even-frequency $s$-wave superfluid case.
\end{abstract}
\maketitle
\par
\section{Introduction}
\par
In cold atom physics, along with the $p$-wave pairing state~\cite{Efremov2002,Regal2004a,Regal2004b,Ticknor2004,Zhang2004,Schunck2005,Gunter2005,Gurarie2005,Ohashi2005,Ho2006,Iskin2006,Gaebler2007,Inada2008,Fuchs2008,Zhang2008,Sato2009,Maier2010,Inotani2012,Nakasuji2013,Liu2014,Inotani2018,Kinnunen2018a,Venu2023} and the Fulde-Ferrell-Larkin-Ovchinnikov state~\cite{Hu2006,He2006,Iskin2006C,Koponen2007,Sheehy2007,Parish2007,Bulgac2008,Chevy2010,Devreese2011,Devreese2011A,Okawauchi2012,Zheng2014,Xu2014,Rosenberg2015,Kinnunen2018,Kawamura2022,Kawamura2023}, the odd-frequency pairing state has recently attracted much attention as a candidate for unconventional Fermi superfluid~\cite{Kalas2008,Arzamasovs2018,Linder2019}. As in the ordinary {\it even}-frequency $s$-wave superfluid state, Cooper pairs are formed also in the odd-frequency superfluid state; however, a crucial difference from the even-frequency case is that the odd-frequency Cooper pair consists of two fermions, not at the same time, but at {\it different} times~\cite{Linder2019,Berezinskii1974}. Since the even-frequency $s$-wave superfluid state has been realized in $^{40}$K~\cite{Regal2004} and $^6$Li~\cite{Zwierlein2004,Kinast2004,Bartenstein2004} Fermi gases by using a Feshbach resonance~\cite{Timmermans2001,Holland2001,Ohashi2002,Chin2010}, exploring an odd-frequency pairing state is an exciting challenge in the current stage of cold Fermi gas physics. 
\par
This unique state was first introduced in the context of superfluid liquid $^3$He by Berezinskii~\cite{Berezinskii1974}, and this idea was later extended to metallic superconductivity~\cite{Balatsky1992,Emery1992,Abrahams1993}. Since then, the odd-frequency superconductivity has extensively been discussed by many researchers in various superconducting systems, such as superconducting junctions~\cite{Asano2007,Yokoyama2007,Yokoyama2011,Tanaka2012,Cayao2020}, strongly correlated electron systems~\cite{Fuseya2003,Shigeta2009,Kusunose2011A,Matsumoto2012,Kusunose2012,Inokuma2024}, Kondo lattice systems~\cite{Tsvelik1993,Tsvelik1994,Hoshino2014A,Hoshino2014B,Funaki2014}, multi-band superconductors~\cite{BlackSchaffer2013,Triola2020}, non-equilibrium systems~\cite{Triola2016,Triola2017,Cayao2021}, as well as superconducting systems with Bogoliubov Fermi surfaces~\cite{Miki2021}. Experimentally, the observation of the paramagnetic Meissner effect, which is considered as a typical odd-frequency superconducting phenomenon near the surface, was recently reported in an Al/Ho/Nb junction~\cite{Bernardo2015A,Bernardo2015B}. More recently, the gapless superconducting density of states, which is considered to be consistent with a bulk odd-frequency pairing state, was also observed through the temperature dependence of the spin-lattice relaxation rate $T_1^{-1}$ in heavy fermion compound CeRh$_{0.5}$Ir$_{0.5}$In$_5$~\cite{Kawasaki2020}. In this way, the odd-frequency superconductivity has been extensively studied both theoretically and experimentally in condensed matter physics. Thus, the realization of an odd-frequency superfluid Fermi atomic gas would also make a great impact on this research field. 
\par
\begin{figure}[t!]
\centering
\includegraphics[width=0.3\textwidth]{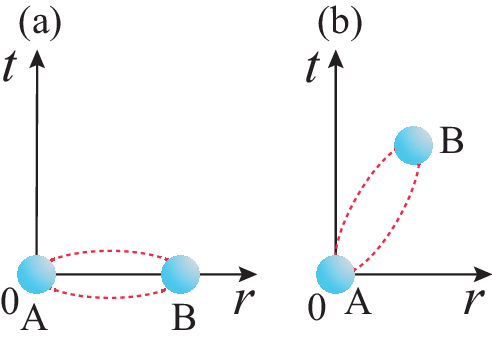}
\caption{Illustration of the space-time structure of a Cooper pair. (a) Even-frequency superfluid state. (b) Odd-frequency superfluid state. $r$ and $t$ are, respectively, the relative coordinate and the relative time between two fermions (``A'' and ``B''). While the equal-time pairing is dominant in the even-frequency Cooper pair, it vanishes in the odd-frequency case.}
\label{fig1}
\end{figure}

As mentioned previously, the unequal-time pairing is characteristic of the odd-frequency superfluid state. Regarding this, we recall that, in the ordinary even-frequency case, the superfluid instability may be interpreted as a kind of Bose-Einstein condensation (BEC) of Cooper-pair `bosons'. Since the even-frequency pairing dominantly occurs between fermions at the same time [see Fig.~\ref{fig1}(a)], this simple picture seems reasonable, especially in the strong-coupling regime, where most fermions form tightly bound molecules. Indeed, such a situation is realized in the BEC regime of the BCS-BEC crossover phenomenon~\cite{NSR1985,Melo1993,Randeria1995,Ohashi2002,Strinati2002,Chen2005,Giorgini2008,Bloch2008,Ohashi2020}, where the superfluid phase transition $T_{\rm c}$ agrees well with the BEC phase transition temperature $T_{\rm BEC}$ of an ideal Bose gas consisting of $N/2$ molecules (where $N$ is the number of fermions). On the other hand, as schematically shown in Fig.~\ref{fig1}(b), the vanishing equal-time pairing in the odd-frequency case is quite different from the naive molecular picture. Thus, this raises the interesting question of whether or not such kind of unequal-time pair still behaves like a boson. 
\par
The purpose of this paper is to theoretically explore the answer to this question. For this purpose, we consider a model odd-frequency superfluid Fermi gas, and take the following steps: (1) At $T=0$, we calculate the space-time structure of the odd-frequency Cooper-pair wavefunction $\varphi_{\rm odd}({\bm r},t)$, within the framework of the mean-field base strong-coupling theory developed by Eagles and Leggett~\cite{Eagles1969,Leggett1980} (which is referred to as the BCS-Eagles-Leggett theory in what follows). We examine the similarity of $\varphi_{\rm odd}({\bm r},t)$ to the even-frequency case, especially in the strong-coupling regime where the even-frequency system is well described by an ideal molecular Bose gas. (2) We evaluate $T_{\rm c}$, by extending the strong-coupling theory for even-frequency $s$-wave Fermi superfluids developed by Nozi\`eres and Schmitt-Rink (NSR)~\cite{NSR1985}, to the odd-frequency case. In the strong-coupling regime where $T_{\rm c}$ in the even-frequency case agrees well with $T_{\rm BEC}$ of an ideal Bose gas, we examine whether or not the same `bosonic picture' is applicable to $T_{\rm c}$ in the odd-frequency case. 
\par
We briefly note that in theoretically treating a bulk odd-frequency Fermi superfluid, the ordinary Hamiltonian formalism is known to give the unphysical result that the superfluid state becomes stable {\it above} $T_{\rm c}$~\cite{Heid1995}. This puzzle was recently solved by employing the path-integral formalism~\cite{Solenov2009,Kusunose2011B}. Following this progress, we also construct our theory by using this approach~\cite{Solenov2009,Kusunose2011B,Fominov2015,note} in this paper. 
\par
This paper is organized as follows: In Sec. II, we explain our formalisms in the superfluid state at $T=0$ based on the BCS-Eagles-Leggett theory~\cite{Eagles1969,Leggett1980}, as well as at $T_{\rm c}$ based on the NSR theory~\cite{NSR1985}. In Sec. III, we show our zero-temperature results on the superfluid order parameter, as well as the Fermi chemical potential. Using these data, we evaluate the pair wavefunction $\varphi_{\rm odd}({\bm r},t)$ in Sec. IV. Here, we also calculate $T_{\rm c}$, to see whether or not a gas of odd-frequency Cooper pairs can be viewed as a Bose gas in the strong-coupling regime. Throughout this paper, we set $\hbar=k_{\rm B}=1$, and the system volume $V$ is taken to be unity, for simplicity.
\par
\par
\section{Formulation}
\subsection{Model odd-frequency Fermi gas}
\par
\par
We consider a single-component Fermi gas with an odd-frequency pairing interaction. Following Refs.~\cite{Solenov2009,Kusunose2011B}, we start from the partition function in the path-integral representation,
\begin{equation}
Z=\int {\cal D}\bar{\psi }{\cal D}\psi e^{-S[\bar{\psi},\psi]}
=\prod _{k}\int d\bar{\psi }_{k} d\psi _{k}
e^{-S[\bar{\psi},\psi]}.
\label{eq.1}
\end{equation}
Here, the Grassmann variable $\displaystyle \psi _{k}$ and its conjugate $\bar{\psi }_{k}$ describe Fermi atoms. In Eq.~(\ref{eq.1}), we have introduced the abbreviated notation $k=({\bm k},i\omega_n)$, where $\omega_n$ is the fermion Matsubara frequency. The action $S[\bar{\psi},\psi]=S_0+S_1$ in Eq.~(\ref{eq.1}) consists of the kinetic term $S_0$ and the pairing interaction term $S_1$, that are given by, respectively,
\begin{equation}
S_0 =\sum _{k}\bar{\psi }_{k}\left[-i\omega_n +\xi _{{\bm k}}\right] \psi _{k},
\label{eq.2}
\end{equation}
\begin{equation}
S_1={1 \over 2\beta}\sum _{q,k,k'} V_{k+q/2,k'+q/2}
{\bar \psi}_{k+q}{\bar \psi}_{-k}\psi_{-k'}\psi_{k'+q}.
\label{eq.3}
\end{equation}
Here $\beta=1/T$, and $q=({\bm q}, i\nu_n)$ with $\nu_n$ being the boson Matsubara frequency. In Eq.~(\ref{eq.2}), $\xi_{\bm k}=\varepsilon_{\bm k}-\mu={\bm k}^2/(2m)-\mu$ is the kinetic energy of a Fermi atom, measured from the Fermi chemical potential $\mu$ (where $m$ is an atomic mass). $V(k,k')=V({\bm k},i\omega_n,{\bm k}',i\omega_n')$ in Eq.~(\ref{eq.3}) is an odd-frequency pairing interaction, where the dependence of the Matsubara frequency ($\omega_n$ and $\omega_n'$) describes retardation effects of this interaction. In this paper, we do not discuss the origin of this interaction, but simply assume the following separable form:
\begin{equation}
V_{k,k'} =-U\gamma ({\bm k}, i\omega_n) \gamma ({\bm k}', i\omega_n').
\label{eq.4}
\end{equation}
Here, $-U~(<0)$ is a coupling strength. (In Appendix A, we briefly explain how this kind of interaction is obtained in the case of phonon-mediated interaction.) When we use Eq.~(\ref{eq.4}), the symmetry of the superfluid order parameter $\Delta({\bm k},i\omega_n)$ is determined by the basis function $\gamma ({\bm k}, i\omega_n)$ as
\begin{equation}
\Delta({\bm k},i\omega_n)=\Delta\gamma({\bm k},i\omega_n),
\label{eq.5}
\end{equation}
where $\Delta$ is a constant. Keeping this in mind, we chose the following {\it odd-frequency} basis function in this paper:
\par
\begin{equation}
\gamma ({\bm k},i\omega_n)
=\frac{\omega_n}{|\omega_n|}
{\sqrt{\omega_n^2 +\xi_{\bm k}^2}
\over
\sqrt{\omega_n^2+\xi_{\bm k}^2+\Lambda^2}
}.
\label{eq.6}
\end{equation}
As shown in Appendix A, $\Lambda$ in Eq.~(\ref{eq.6}) is related to the frequency of the Einstein phonon in a phonon-mediated interaction. In this sense, $\Lambda$ may be regarded as a parameter to tune retardation effects coming from the model interaction in Eq.~(\ref{eq.4}).
\par
\begin{figure}[t!]
\centering
\includegraphics[width=0.3\textwidth]{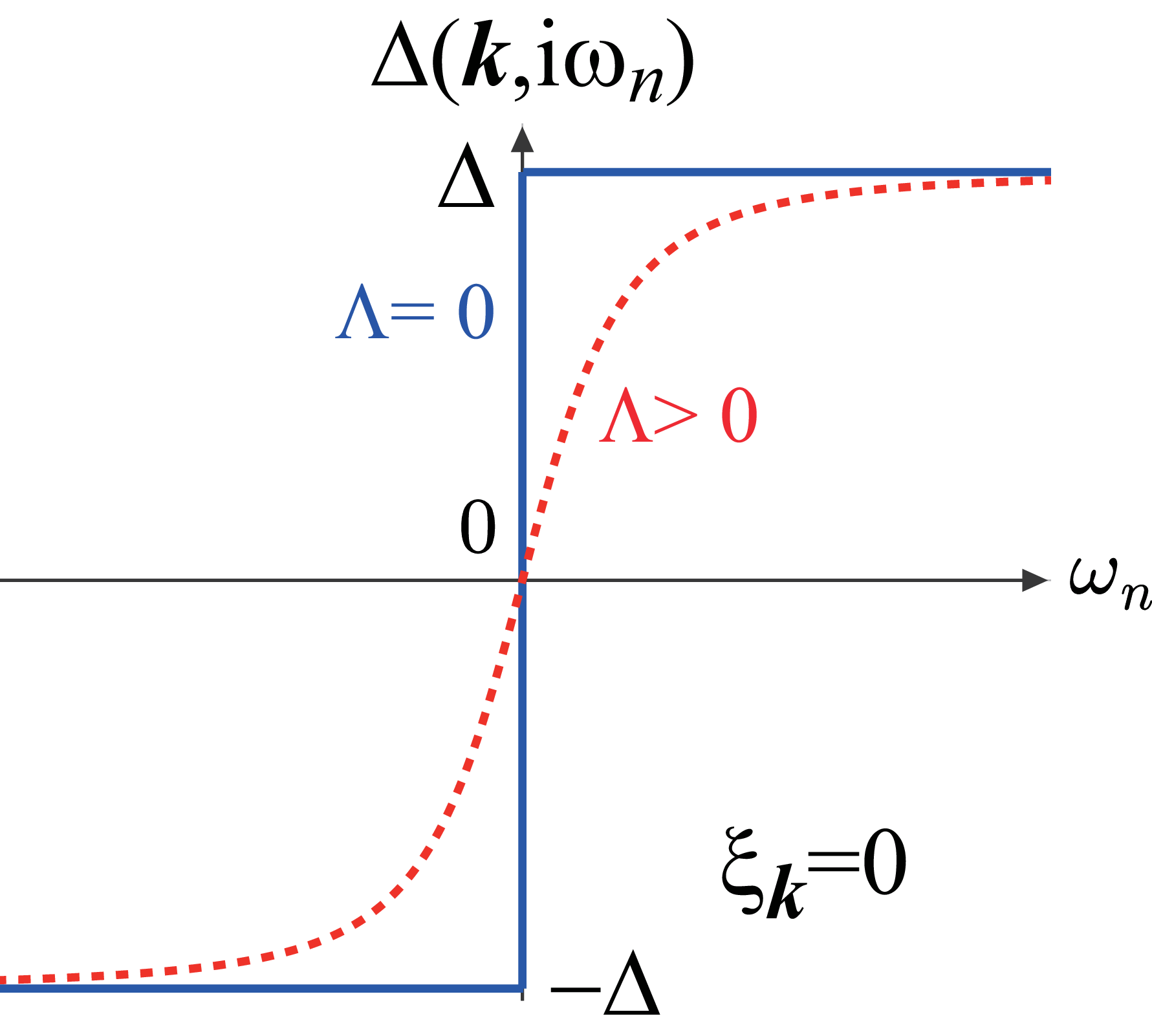}
\caption{Illustration of the odd-frequency superfluid order parameter $\Delta({\bm k},i\omega_n)$ in Eq.~(\ref{eq.5}) as a function of $\omega_n$, when $\xi_{\bm k}=0$.}
\label{fig2}
\end{figure}

The resulting superfluid order parameter $\Delta({\bm k},i\omega_n)$ in Eq.~(\ref{eq.5}) has the {\it odd-frequency spin-triplet $s$-wave} pairing symmetry~\cite{note2}. That is, $\Delta({\bm k},i\omega_n)$ is an odd function with respect to $\omega_n$ (see Fig.~\ref{fig2}), and isotropic in momentum space.
\par
\par
Here, we comment on the reason for the choice of Eq.~(\ref{eq.6}): In superconductivity literature, the separable pairing interaction in Eq.~(\ref{eq.4}) is frequently used to describe an anisotropic (unconventional) superconducting order parameter in momentum space. (In this case, the frequency dependence of the basis function is usually dropped.) For example, the $d_{x^2-y^2}$-wave superconductivity discussed in high-$T_{\rm c}$ cuprates can be described by setting $\gamma({\bm k})\propto k_x^2-k_y^2$. In this way, the basis function can flexibly be chosen in terms of the momentum dependence of the pairing state. On the other hand, one needs to be careful in the odd-frequency case. For example, one may think that the simpler basis function $\gamma({\bm k},i\omega_n)=\omega_n$ is more tractable than Eq.~(\ref{eq.6}) as a model odd-frequency pairing interaction. However, the diagonal component of the resulting mean-field BCS Green's function~\cite{Linder2019},
\begin{equation}
G_{\rm odd}^{(1,1)}({\bm k},i\omega_n)=
-
{i\omega_n+\xi_{\bm k} \over [1+|\Delta|^2]\omega_n^2+\xi_{\bm k}^2},
\label{eq.7}
\end{equation}
gives the following single-particle spectrum weight $A({\bm k},\omega)$:
\begin{eqnarray}
A({\bm k},\omega)
&=&
-{1 \over \pi}
{\rm Im}
\left[
G_{\rm odd}^{(1,1)}({\bm k},i\omega_n\to\omega+i\delta)
\right]
\nonumber\\
&=&
{1+\sqrt{1+|\Delta|^2} \over 2[1+|\Delta|^2]}\delta(\omega-\xi_{\bm k}) \nonumber\\ 
&+&
{1-\sqrt{1+|\Delta|^2} \over 2[1+|\Delta|^2]}\delta(\omega+\xi_{\bm k}),
\label{eq.8}
\end{eqnarray}
where $\delta$ in the first line is an infinitesimally small positive number. Equation~(\ref{eq.8}) is unphysical, because the last term is negative. In contrast, we will show in Sec. III that Eq.~(\ref{eq.6}) gives positive $A({\bm k},\omega)$~\cite{alt}. (We show in Appendix A that the positivity of $A({\bm k},\omega)$ again breaks down when we drop the factor $\xi_{\bm k}^2$ in Eq.~(\ref{eq.6})). 
\par
We introduce the Cooper-pair Bose field $\Phi$ as well as its conjugate field ${\bar \Phi}$, by way of the Stratonovich-Hubbard transformation~\cite{Stratonovich1958,Hubbard1959}. Executing the fermion path-integrals, we have (for the derivation, see Appendix B)
\begin{eqnarray}
Z
&\propto& 
\int {\cal D}{\bar \Phi}{\cal D}\Phi 
e^{-S_{\rm eff}[{\bar \Phi}, \Phi]}
\nonumber
\\
&=&
\prod_q \int d{\rm Re}[\Phi_q] d{\rm Im}[\Phi_q]
e^{-S_{\rm eff}[{\bar \Phi}, \Phi]},
\label{eq.18}
\end{eqnarray}
where the effective action $S_{\rm eff}$ is given by
\begin{equation}
S_{\rm eff}=
-{1 \over 2}{\rm Tr}\ln\left[-{\hat G}^{-1}\right]
+
\sum_q
{{\bar \Phi}_q \Phi_q \over 2U} 
+{\beta \over 2}\sum_{\bm k}\xi_{\bm k}.
\label{eq.19}
\end{equation}
Here,
\begin{align}
&{\hat G}^{-1}_{kk'}=
\nonumber
\\
&\left(
\begin{array}{cc}
\displaystyle
[i\omega_n-\xi_{\bm k}]\delta _{k,k'} & 
\frac{1}{\sqrt{\beta}}\gamma(\frac{\bm{k}+\bm{k'}}{2},\frac{i\omega_n+i\omega_n'}{2}) \Phi_{k-k'} \\
\frac{1}{\sqrt{\beta}}\gamma(\frac{\bm{k}+\bm{k'}}{2},\frac{i\omega_n+i\omega_n'}{2}){\bar \Phi}_{k'-k} & 
[i\omega_n+\xi_{\bm k}]\delta_{k,k'}
\end{array}
\right)
\label{eq.17}
\end{align}
is the inverse of the $2\times 2$ matrix single-particle thermal Green's function~\cite{Iskin2005,Iskin2006A,Iskin2006B,Cao2013,Tempere2008}.
\par
\par
\subsection{Odd-frequency superfluid state at $T=0$}
\par
Eagles and Leggett pointed out that the mean-field BCS theory, which was originally invented for weak-coupling superconductivity, is actually applicable to the whole BCS-BEC crossover region (at least qualitatively), when we deal with the BCS gap equation, together with the equation for the number $N$ of fermions, to self-consistently determine the superfluid order parameter and the Fermi chemical potential. In this paper, we extend this scheme to the odd-frequency Fermi superfluid described by the effective action $S_{\rm eff}$ in Eq.~(\ref{eq.19}).
\par
It is well-known that the mean-field BCS theory corresponds to the saddle-point approximation in the path-integral formalism. In this approximation, the path-integrals with respect to the Cooper-pair fields ${\bar \Phi}$ and $\Phi$ in Eq.~(\ref{eq.18}) are replaced by the representative value at the saddle-point solution~\cite{Tempere2012}. The resulting partition function ($\equiv Z_{\rm SP}$) has the form,
\begin{equation}
Z_{\rm SP} = 
e^{-\left[
{\beta\Delta^* \Delta \over 2U}
+
{\beta \over 2}\sum_{\bm k}\xi_{\bm k}
-{1 \over 2}\ln
\left(-\det\left[-{\hat G}_{\rm SP}^{-1}\right]\right)\right]
} ,
\label{eq.20}
\end{equation}
where $[{\hat G}_{\rm SP}^{-1}]_{kk'}={\hat G}_{\rm odd}(k)^{-1}\delta_{k,k'}$ with 
\begin{equation}
{\hat G}_{\rm odd}(k)=
{1 \over 
i\omega_n-\xi_{\bm k}\tau_3+
\begin{pmatrix}
0 & 
\Delta \gamma({\bm k},i\omega_n) \\
\Delta^* \gamma({\bm k},i\omega_n) &
0
\end{pmatrix}
}
\label{eq.21}
\end{equation}
being the $2\times 2$ matrix mean-field single-particle thermal Green's function in the odd-frequency superfluid state (where $\tau_{i=1,2,3}$ are the Pauli matrices acting on particle-hole space). In obtaining $Z_{\rm SP}$ in Eq.~(\ref{eq.20}), we have chosen the Cooper-pair fields at the saddle point as~\cite{Solenov2009,Kusunose2011B}
\begin{eqnarray}
\left\{
\begin{array}{l}
\Phi_q =\sqrt{\beta } \Delta \delta_{q,0},\\
{\bar \Phi}_q =\sqrt{\beta }\Delta^* \delta_{q,0}.
\end{array}
\right.
\label{eq.22}
\end{eqnarray}
The mean-field superfluid order parameter $\Delta$ in Eq.~(\ref{eq.22}) is determined from the saddle point condition,
\begin{equation}
{\partial \Omega_{\rm MF} \over \partial \Delta^*}=0.
\label{eq.23}
\end{equation}
Here,
\begin{align}
\Omega_{\rm MF}
&=
-T\ln Z_{\rm SP}
\nonumber
\\
&=
{|\Delta|^2 \over 2U}
+{1 \over 2}\sum_{\bm k}\xi_{\bm k}
-{1 \over 2\beta}\sum_k \nonumber\\
&\times\ln\left[\omega_n^2+\xi_{\bm k}^2+|\Delta|^2\gamma({\bm k},i\omega_n)^2\right]
\label{eq.24}
\end{align} 
is just the mean-field thermodynamic potential. Substituting Eq.~(\ref{eq.24}) into Eq.~(\ref{eq.23}), we obtain the BCS-type gap equation,
\begin{eqnarray}
1
&=&
{U \over \beta}
\sum_{{\bm k},\omega_n}
{\gamma({\bm k},i\omega_n)^2 \over
\omega_n^2+\xi_{\bm k}^2+|\Delta|^2\gamma({\bm k},i\omega_n)^2}
\nonumber
\\
&=&U\sum _{\bm k}
{1 \over 2{\cal E}_{\rm odd}({\bm k},\Lambda)}
\tanh\left({{\cal E}_{\rm odd}({\bm k},\Lambda) \over 2T}\right),
\label{eq.25}
\end{eqnarray}
where
\begin{equation}
{\cal E}_{\rm odd}({\bm k},\Lambda) =\sqrt{\xi_{\bm k}^2+\Lambda^2+|\Delta|^{2}}
\label{eq.26}
\end{equation}
describes Bogoliubov single-particle excitations. We briefly note that the gap equation~(\ref{eq.25}) can also be obtained from the (1,2) component $G_{\rm odd}^{(1,2)}$ of the $2\times 2$ matrix Green's function in Eq.~(\ref{eq.21}) as
\begin{equation}
\Delta\gamma({\bm k},i\omega_n)=
{1 \over \beta}\sum_{k'}V_{k,k'}G_{\rm odd}^{(1,2)}(k').
\label{eq.27}
\end{equation}
\par
As pointed out in Refs.~\cite{Solenov2009,Kusunose2011B}, the choice in Eq.~(\ref{eq.22}) guarantees the expected thermodynamic behavior that the superfluid phase becomes stable {\it below} $T_{\rm c}$. In contrast, the Hamiltonian formalism gives the opposite result that the second equation in Eq.~(\ref{eq.22}) is replaced by ${\bar \Phi}_q=-\sqrt{\beta}\Delta^*\delta_{q,0}$~\cite{Solenov2009,Kusunose2011B}, which lead to the unphysical situation, as mentioned previously.
\par
We remove the ultraviolet divergence involved in the gap equation~(\ref{eq.25}) by the same prescription as that used in the BCS-BEC crossover theory for even-frequency $s$-wave Fermi superfluids~\cite{Melo1993,Randeria1995,Leggett1980}: We measure the interaction strength in terms of the $s$-wave scattering length $a_s$ in an assumed two-component Fermi gas with a contact-type $s$-wave pairing interaction $H_{\rm I}\equiv-U\delta({\bm r}_1-{\bm r}_2)$, which is related to $-U$ as~\cite{Randeria1995},
\begin{equation}
{4\pi a_s \over m}
=-
{U \over 1-U\sum_{\bm k}{1 \over 2\varepsilon_{\bm k}}}.
\label{eq.28}
\end{equation}
We then rewrite the gap equation~(\ref{eq.25}) as
\begin{equation}
1=
-{4\pi a_s \over m}
\sum_{\bm k}
\left[
{1 \over 2{\cal E}_{\rm odd}({\bm k},\Lambda)}
\tanh
\left({{\cal E}_{\rm odd}({\bm k},\Lambda) \over 2T}\right)
-
{1 \over 2\varepsilon_{\bm k}}
\right].
\label{eq.29}
\end{equation}
In this scale, the weak-coupling (strong-coupling) regime is described as $(k_{\rm F}a_s)^{-1}\lesssim -1$ [$(k_{\rm F}a_s)^{-1}\gesim +1$], where $k_{\rm F}$ is the Fermi momentum.
\par
In the BCS-Eagles-Leggett scheme~\cite{Eagles1969,Leggett1980}, we solve the gap equation~(\ref{eq.29}) at $T=0$, together with the equation for the number $N$ of Fermi atoms. The latter equation is obtained from the $(1,1)$ component of the Green's function in Eq.~(\ref{eq.21})~\cite{note22}: 
\begin{align}
N
=&
{1 \over \beta}\sum_k G_{\rm odd}^{(1,1)}(k)\Bigr|_{T=0}
\nonumber
\\
=&
{\Lambda^2 \over |\Delta|^2+\Lambda^2}
\sum_{\bm k}\Theta(-\xi_{\bm k})
\nonumber
\\
&+
{|\Delta|^2 \over |\Delta|^2+\Lambda^2}
\sum_{\bm k}
{1 \over 2}
\left[
1-{\xi_{\bm k} \over {\cal E}_{\rm odd}({\bm k},\Lambda)}
\right],
\label{eq.30}
\end{align}
where $\Theta(-\xi_{\bm k})$ is the step function.
\par
\par
\subsection{Pair wavefunction}
\par
To examine the space-time structure of the odd-frequency Cooper pair, we consider the pair wavefunction at $T=0$, given by~\cite{Kadin2007}
\begin{eqnarray}
\varphi_{\rm odd}({\bm r},t)
&=&
\langle\psi({\bm r},t)\psi(0,0)\rangle
\nonumber
\\
&=&
-i\sum_{\bm k}e^{-i{\bm k}\cdot{\bm r}}
{\cal G}_{\rm odd}^{<,(1,2)}({\bm k},-t),
\label{eq.x1}
\end{eqnarray}
where the lesser Green's function ${\cal G}_{\rm odd}^{<,(1,2)}({\bm k},t)$ is related to the (1,2)-component of the thermal Green's function ${\hat G}({\bm k},i\omega_n)$ in Eq.~(\ref{eq.21}) as~\cite{Rammer}
\begin{equation}
{\cal G}_{\rm odd}^{<,(1,2)}({\bm k},t)=
\int_{-\infty}^\infty{d\omega \over 2\pi}
e^{-i\omega t}
{\cal G}_{\rm odd}^{<,(1,2)}({\bm k},\omega),
\label{eq.x0}
\end{equation}
\begin{align}
{\cal G}_{\rm odd}^{<,(1,2)}({\bm k},\omega)
=&-f(\omega)
\left[
G_{\rm odd}^{(1,2)}({\bm k},i\omega_n\to\omega+i\delta)|_{\omega_n>0}
\right.
\nonumber
\\
&-
\left.
G_{\rm odd}^{(1,2)}({\bm k},i\omega_n\to\omega-i\delta)|_{\omega_n<0}
\right].
\label{eq.x2}
\end{align}
Here, $f(\omega)$ is the Fermi distribution function, which equals the step function $\Theta(-\omega)$ at $T=0$. 
\par
To grasp the background physics of $\varphi_{\rm odd}({\bm r},t)$, it is helpful to recall the pair wavefunction discussed in the ordinary (even-frequency) spin-singlet $s$-wave pairing state given by~\cite{Kadin2007},
\begin{eqnarray}
\varphi_{\rm even}({\bm r})
&=&
\langle\psi_\downarrow({\bm r})\psi_\uparrow(0)\rangle
\nonumber
\\
&=&
{\Delta_{\rm even} \over 4\pi^2 r}\int_0^\infty dk 
{k \over {\cal E}_{\rm even}({\bm k})}
\sin(kr).
\label{eq.y1}
\end{eqnarray}
The outline of the derivation is explained in Appendix C. In Eq.~(\ref{eq.y1}), the field operator $\psi_{\sigma=\uparrow,\downarrow}({\bm r})$ describes fermions with pseudospin $\sigma=\uparrow,\downarrow$, and ${\cal E}_{\rm even}({\bm k})=\sqrt{\xi_{\rm even}({\bm k})^2+\Delta_{\rm even}^2}$, where $\xi_{\rm even}({\bm k})=\varepsilon_{\bm k}-\mu_{\rm even}$ is the kinetic energy, measured from the Fermi chemical potential $\mu_{\rm even}$. The $s$-wave superfluid order parameter $\Delta_{\rm even}$ (which is taken to be real, for simplicity), as well as $\mu_{\rm even}$, are determined from the BCS-Eagles-Leggett coupled equations~\cite{Leggett1980} 
\begin{equation}
1=-{4\pi a_s \over m}\sum_{\bm k}
\left[
{1 \over 2{\cal E}_{\rm even}({\bm k})}
-{1 \over 2\varepsilon_{\bm k}}
\right],
\label{eq.x4b}
\end{equation}
\begin{equation}
N=\sum_{\bm k}
\left[
1-{\xi_{\rm even}({\bm k}) \over {\cal E}_{\rm even}({\bm k})}
\right].
\label{eq.x4c}
\end{equation}
\par
Deep inside the strong-coupling regime, where $\mu_{\rm even}\simeq -1/(2ma_s^2)<0$ and $|\mu_{\rm even}|\gg\Delta_{\rm even}$~\cite{Leggett1980,Randeria1995}, one finds
\begin{equation}
\varphi_{\rm even}({\bm r})\simeq
{m\Delta_{\rm even} \over 4\pi r}e^{-r/a_s}.
\label{eq.x6}
\end{equation}
Apart from the unimportant constant factor, Eq.~(\ref{eq.x6}) is just the same form as the wavefunction of a two-body bound state with the binding energy~\cite{Randeria1995,Ohashi2020},
\begin{equation}
E_{\rm bind}={1 \over ma_s^2}.
\label{eq.xx6}
\end{equation}
\par
In the weak-coupling BCS regime (where $\mu_{\rm even}\simeq\varepsilon_{\rm F}$, with $\varepsilon_{\rm F}$ being the Fermi energy), Eq.~(\ref{eq.y1}) is reduced to
\begin{equation}
\varphi_{\rm even}({\bm r})\simeq
{m\Delta_{\rm even} \over 2\pi^2 r}
K_0\left({r \over \xi_{\rm coh}}\right)\sin(k_{\rm F} r),
\label{eq.x7}
\end{equation}
where $K_0(x)$ is the zeroth modified Bessel function, and $\xi_{\rm coh}=v_{\rm F}/\Delta_{\rm even}$ is the BCS coherence length~\cite{deGennes} (where $v_{\rm F}$ is the Fermi velocity). Noting that $K_0(x)\simeq \sqrt{\pi/(2x)}{\rm exp}(-x)$ for $x\gg 1$, one finds that Eq.~(\ref{eq.x7}) physically describes a Cooper pair whose spatial size is comparable to the coherence length $\xi_{\rm coh}$. In this sense, $\varphi_{\rm odd}({\bm r},t)$ in Eq.~(\ref{eq.x1}) is a natural extension of the equal-time pair wavefunction, to include effects of time difference between two fermions involved in the Cooper pair.
\par
\begin{figure}[t]
\centering
\includegraphics[width=0.4\textwidth]{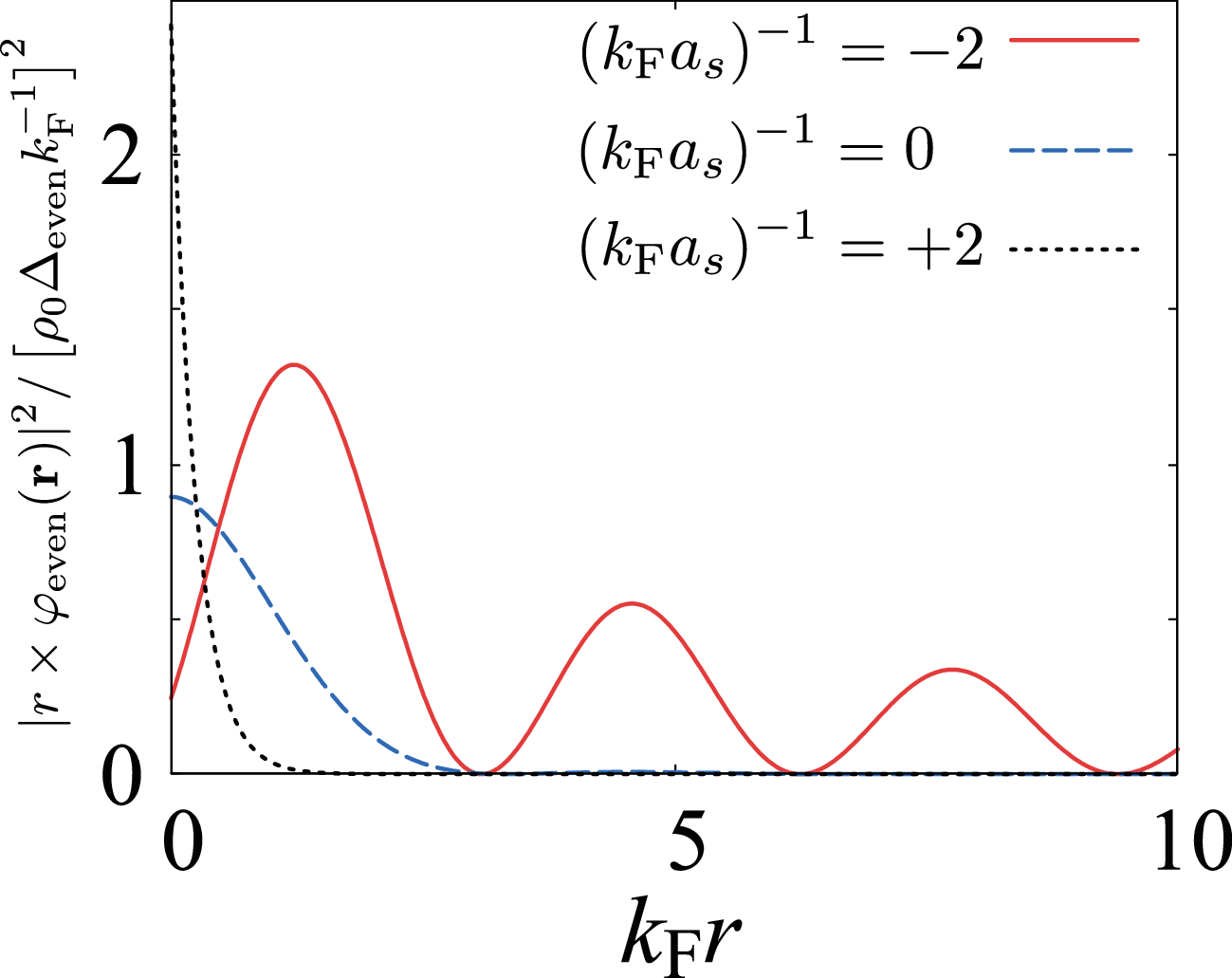}
\caption{Spatial variation of equal-time pair wavefunction $\varphi_{\rm even}({\bm r})$ in Eq.~(\ref{eq.x3}) at $T=0$. $\rho_0$ is given in Eq.~(\ref{eq.xxx1}).
}
\label{fig3}
\end{figure}
\par
Since the (1,2)-component $G_{\rm odd}^{(1,2)}({\bm k},i\omega_n)$ of the odd-frequency Green's function in Eq.~(\ref{eq.21}) is isotropic in momentum space, one can rewrite $\varphi_{\rm odd}({\bm r},t)$ in Eq.~(\ref{eq.x1}) as
\begin{align}
\varphi_{\rm odd}({\bm r},t)=
&{i \over 2\pi^2 r}
\int_0^\infty kdk \sin(kr)
\int_{-\infty}^\infty{d\omega \over 2\pi}
\left[
e^{i\omega t}f(\omega)
\right.
\nonumber
\\
&
+\left.e^{-i\omega t}f(-\omega)
\right]
G_{\rm odd}^{(1,2)}({\bm k},i\omega_n\to\omega+i\delta)|_{\omega_n>0}.
\label{eq.x8}
\end{align}
In obtaining Eq.~(\ref{eq.x8}), we have used the symmetry property,
\begin{align}
G_{\rm odd}^{(1,2)}&({\bm k},i\omega_n\to-\omega-i\delta)|_{\omega_n<0}
\nonumber
\\
=&-
G_{\rm odd}^{(1,2)}({\bm k},i\omega_n\to\omega+i\delta)|_{\omega_n>0}.
\label{eq.x9}
\end{align}
At $t=0$, because the retarded Green's function $G_{\rm odd}^{(1,2)}({\bm k},i\omega_n\to\omega+i\delta)$ is analytic in the upper-half complex plane, the $\omega$ integration in Eq.~(\ref{eq.x8}) vanishes by closing the integral path in the upper-half plane. Thus, while the equal-time pair wavefunction $\varphi_{\rm even}({\bm r})$ is non-zero in the even-frequency case as shown in Fig.~\ref{fig3}, one finds,
\begin{equation}
\varphi_{\rm odd}({\bm r},t=0)=0,
\label{eq.x9b}
\end{equation}
in the odd-frequency case.
\par
For later convenience, we introduce the time-dependent even-frequency pair wavefunction as,
\begin{align}
\varphi_{\rm even}({\bm r},t)
&=
\langle\psi_\downarrow({\bm r},t)\psi_\uparrow(0,0)\rangle
\nonumber
\\ 
&=
{\Delta_{\rm even} \over 4\pi^2 r}\int_0^\infty dk 
{k \over {\cal E}_{\rm even}({\bm k})}
\sin(kr)e^{-i{\cal E}_{\rm even}({\bm k})t}.
\label{eq.y2}
\end{align}
We summarize the derivation of Eq.~(\ref{eq.y2}) in Appendix C.
\par
\par
\subsection{Superfluid phase transition temperature $T_{\rm c}$}
\par
We next consider $T_{\rm c}$ within the framework of the NSR theory~\cite{NSR1985}. In the path-integral formalism, this strong-coupling theory corresponds to the Gaussian fluctuation theory with respect to the auxiliary Bose fields $\Phi_q$ and ${\bar \Phi}_q$ around $\Phi_q={\bar \Phi}_q=0$~\cite{Melo1993,Randeria1995}. Expanding the effective action $S_{\rm eff}$ in Eq.~(\ref{eq.19}) with respect to these fields up to the second order, one obtains the NSR partition function $Z_{\rm NSR}$ as,
\begin{align}
Z_{\rm NSR}
&=
e^{-{\beta \over 2}\sum_{\bm k}\xi_{\bm k}+{1 \over 2}\sum_k\ln(\omega_n^2+\xi_{\bm k}^2)}
\int{\cal D}{\bar \Phi}{\cal D}\Phi
e^{{1 \over 2U}
{\bar \Phi}_q
[1-U\Pi(q)]
\Phi_q}
\nonumber
\\
&=
e^{-{\beta \over 2}\sum_{\bm k}\xi_{\bm k}+{1 \over 2}\sum_k\ln(\omega_n^2+\xi_{\bm k}^2)}
\prod_q{1 \over 1-U\Pi(q)},
\label{eq.31}
\end{align}
where we have dropped an unimportant constant factor. In Eq.~(\ref{eq.31}),
\begin{align}
\Pi(q)
&=
{1 \over \beta}\sum_k
\gamma\left({\bm k},i\omega_n+i\nu_n/2\right)^2
\nonumber
\\
&\times G_0\left({\bm k}+{{\bm q} \over 2},i\omega_n+i\nu_n\right)
G_0\left(-{\bm k}+{{\bm q} \over 2},-i\omega_n\right)
\nonumber
\\ 
&=
-\sum_{\bm k}
{1-f\left(\xi_{-{\bm k}+{\bm q}/2}\right)-f\left(\xi_{{\bm k}+{\bm q}/2}\right)
\over 
i \nu_n-\xi_{-{\bm k}+{\bm q}/2}-\xi_{{\bm k}+{\bm q}/2}}
\nonumber
\\
&+\sum_{\bm k}{1 \over i \nu_n-\xi_{-{\bm k}+{\bm q}/2}-\xi_{{\bm k}+{\bm q}/2}} \sum_{\sigma, \sigma'=\pm 1}
{\sigma' \Lambda^2 \over 2\sqrt{\xi_{\bm k}^2+\Lambda^2}}
\nonumber
\\
&\times
{f\left(\xi_{\sigma{\bm k}+{\bm q}/2}\right)-f\left(\sigma'\sqrt{\xi_{\bm k}^2+\Lambda^2}+i\nu_n/2\right)
\over 
\xi_{\sigma{\bm k}+{\bm q}/2}-\sigma'\sqrt{\xi_{\bm k}^2+\Lambda^2}-i\nu_n/2}
\label{eq.32}
\end{align}
is the pair-correlation function describing fluctuations in the Cooper channel, with $G_0^{-1}({\bm k},i\omega_n)=i\omega_n-\xi_{\bm k}$ being the free single-particle Green's function. The resulting thermodynamic potential $\Omega_{\rm NSR}\equiv -T\ln Z_{\rm NSR}$ has the form,
\begin{equation}
\Omega_{\rm NSR}=
{1 \over 2}\sum_{\bm k}\xi_{\bm k}-{1 \over 2\beta}
\sum_k\ln(\omega_n^2+\xi^2_{\bm k})
+{1 \over \beta}\sum_q
\ln[1-U\Pi(q)].
\label{eq.33}
\end{equation}
The NSR number equation is then obtained from the thermodynamic identity $N=-\partial\Omega_{\rm NSR}/\partial\mu$, which gives
\begin{equation}
N=\sum_{\bm k}f(\xi_{\bm k})
-{1 \over \beta}\sum_q{\partial \over \partial\mu}
\ln
\left[
1+{4\pi a_s \over m}
\left[\Pi(q)-\sum_{\bm k}{1 \over 2\varepsilon_{\bm k}}
\right]
\right],
\label{eq.34}
\end{equation}
where we have removed the ultraviolet divergence involved in $\Pi(q)$, by replacing the bare interaction $U$ with the $s$-wave scattering length $a_s$ in Eq.~(\ref{eq.28}).
\par
We solve the $T_{\rm c}$ equation,
\begin{equation}
1=-{4\pi a_s \over m}
\sum_{\bm k}
\left[
{1 \over 2\sqrt{\xi_{\bm k}^2+\Lambda^2}}
\tanh
\left(
{\sqrt{\xi_{\bm k}^2+\Lambda^2} \over 2T}
\right)
-{1 \over 2\varepsilon_{\bm k}}
\right]
\label{eq.35}
\end{equation}
[which is obtained from the gap equation~(\ref{eq.29}) with $\Delta=0$], together with the NSR number equation~(\ref{eq.34}), to consistently determine $T_{\rm c}$ and $\mu(T_{\rm c})$.
\par
\begin{figure}[t]
\centering
\includegraphics[width=0.4\textwidth]{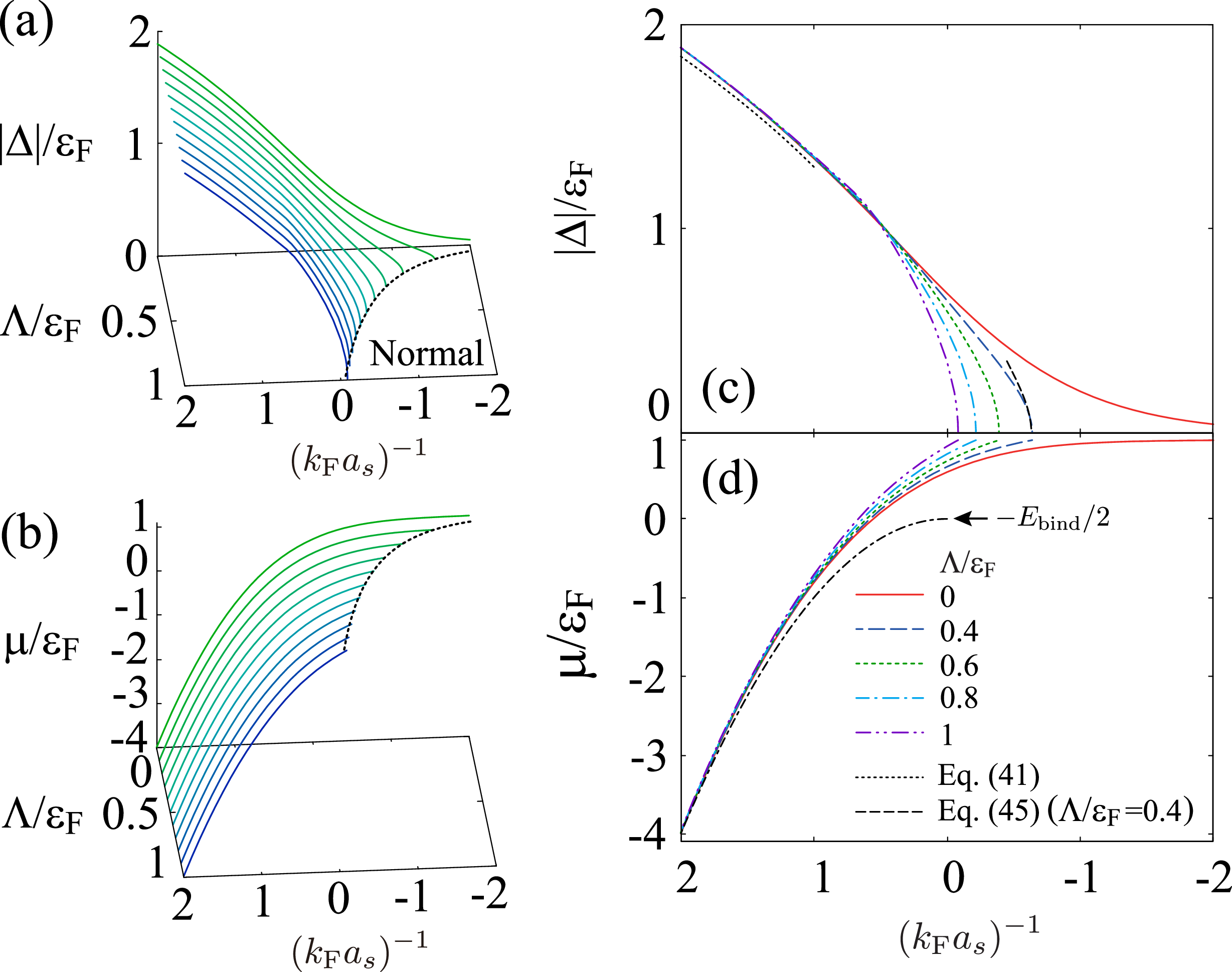}
\caption{Calculated (a) magnitude $|\Delta|$ of the odd-frequency superfluid order parameter and (b) the Fermi chemical potential $\mu$ at $T=0$. Panels (c) and (d), respectively, show $|\Delta|$ and $\mu$ as functions of the interaction strength. In panel (d), $E_{\rm bind}$ is the binding energy of a two-body bound state given in Eq.~(\ref{eq.xx6}). We also plot the approximate result given in Eqs.~(\ref{eq.36}) and~(\ref{eq.39}) in panel (c).}
\label{fig4}
\end{figure}
\par
\section{Ground state properties of odd-frequency Fermi superfluid}
\par
In this section, we consider the odd-frequency Fermi superfluid state at $T=0$, within the framework of the BCS-Eagles-Leggett theory explained in Sec. II.B.
\par
Figures~\ref{fig4}(a) and (b), respectively, show the odd-frequency superfluid order parameter $|\Delta|$ and the Fermi chemical potential $\mu$. When $\Lambda=0$, because the basis function is reduced to $\gamma({\bm k},i\omega_n)={\rm sgn}(\omega_n)$, both the gap equation~(\ref{eq.29}) and the number equation~(\ref{eq.30}) have the same form as Eqs.~(\ref{eq.x4b}) and~(\ref{eq.x4c}), respectively, for the even-frequency $s$-wave superfluid state. [Note that the basis function only appears as $\gamma({\bm k},i\omega_n)^2$ in Eqs.~(\ref{eq.29}) and~(\ref{eq.30}).] The resulting $|\Delta|$ and $\mu$ thus exhibit the same behavior as in the even-frequency $s$-wave case. That is, the system is always in the superfluid state, irrespective of the interaction strength. Particularly in the strong-coupling regime [$(k_{\rm F}a_s)^{-1}\gesim +1$], the magnitude of the superfluid order parameter approaches~\cite{Ohashi2020} [see Fig.~\ref{fig4}(c)]
\begin{equation}
|\Delta|=\varepsilon_{\rm F}\sqrt{16 \over 3\pi k_{\rm F}a_s}.
\label{eq.36}
\end{equation}
In addition, as shown in Fig.~\ref{fig4}(d), the Fermi chemical potential $\mu$ in this regime becomes negative, to approach
\begin{equation}
\mu=-{1 \over 2}E_{\rm bind},
\label{eq.37}
\end{equation}
where $E_{\rm bind}$ is the binding energy of a two-body bound state given in Eq.~(\ref{eq.xx6}). These results are just the same as those in the even-frequency $s$-wave superfluid Fermi gas in the BCS-BEC crossover region~\cite{NSR1985,Melo1993,Randeria1995,Ohashi2002,Strinati2002,Chen2005,Giorgini2008,Bloch2008,Ohashi2020}.
\par
When $\Lambda>0$, on the other hand, we see in Fig.~\ref{fig4}(a) that the superfluid state vanishes, when the pairing interaction becomes weak to some extent. This is because of the suppression of the present odd-frequency pairing interaction in the low-energy region. Indeed, when $\Lambda>0$, the factor $\gamma({\bm k},i\omega_n)^2$ in the numerator in the first line in the gap equation~(\ref{eq.25}) becomes small for small $|\omega_n|$, which physically means the weakening of the pairing interaction around the Fermi level. In the weak-coupling regime at $T=0$ (where $\mu\simeq\varepsilon_{\rm F}$ and $|\Delta|\ll\varepsilon_{\rm F}$), we approximate the second line in Eq.~(\ref{eq.25}) to
\begin{eqnarray}
1
&\simeq& U\rho_0
\int_0^{\omega_{\rm c}}
d\xi {1 \over \sqrt{\xi^2+\Lambda^2+|\Delta|^2}}
\nonumber
\\
&\simeq&
U\rho_0\ln
\left(
{2\omega_{\rm c} \over \sqrt{\Lambda^2+|\Delta|^2}}
\right),
\label{eq.38}
\end{eqnarray}
where $\omega_{\rm c}~(\gg\Lambda)$ is an energy cutoff, and  
\begin{equation}
    \rho_0=\frac{mk_{\rm F}}{2\pi^2}
    \label{eq.xxx1}
\end{equation}
is the density of states in a single-component free Fermi gas at the Fermi level. We then immediately find that the interaction strength needs to exceed the threshold value $U_{\rm c}\rho_0\equiv (\ln(2\omega_{\rm c}/\Lambda))^{-1}$, in order to obtain a non-zero $\Delta$. More quantitatively, applying the same discussion to the renormalized gap equation~(\ref{eq.29}) at $T=0$, one obtains
\begin{equation}
|\Delta|=\varepsilon_{\rm F}
\sqrt{
\left(
{8 \over e^2} e^{{\pi \over 2}{1 \over k_{\rm F}a_s}}
\right)^2
-\left({\Lambda \over \varepsilon_{\rm F}}\right)^2
},
\label{eq.39}
\end{equation}
which gives the threshold interaction strength $(k_{\rm F}a_s^{\rm c})^{-1}$ as,
\begin{equation}
(k_{\rm F}a_s^{\rm c})^{-1}=
{2 \over \pi}
\ln
\left(
{e^2 \over 8}{\Lambda \over \varepsilon_{\rm F}}
\right),
\label{eq.40}
\end{equation}
As seen in Fig.~\ref{fig4}(c), Eq.~(\ref{eq.39}) well describes the behavior of $|\Delta|$ around $(k_{\rm F}a_s^{\rm c})^{-1}$ given in Eq.~(\ref{eq.40}).
\par
\begin{figure}[t]
\centering
\includegraphics[width=0.45\textwidth]{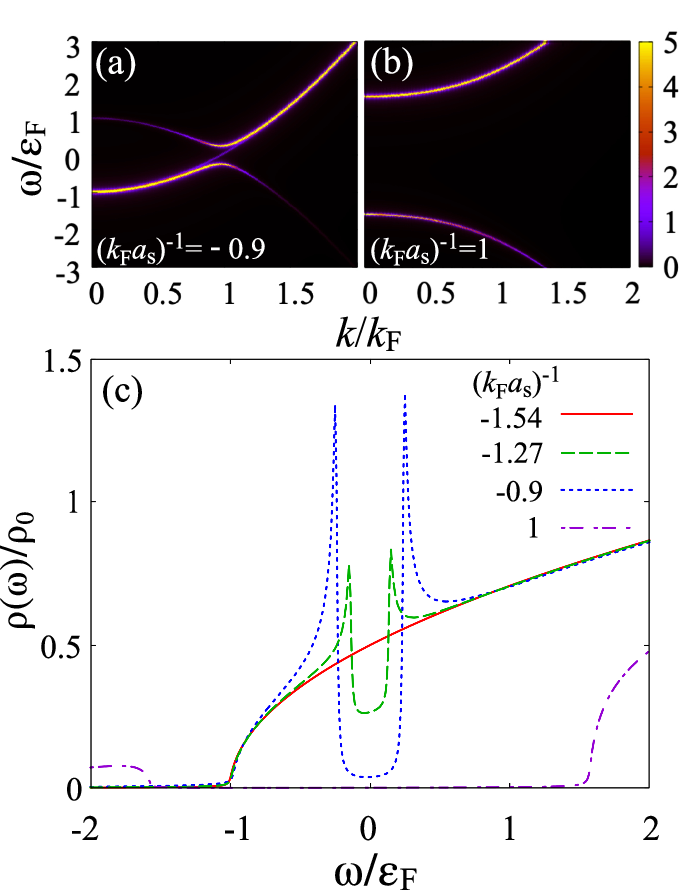}
\caption{Calculated intensity of single-particle spectral weight $A({\bm k},\omega)$ in the odd-frequency superfluid state at $T=0$  and $\Lambda/\varepsilon_{\rm F}=0.1$. (a) Weak-coupling regime [$(k_{\rm F}a_s)^{-1}=-0.9$]. (b) Strong-coupling regime [($k_{\rm F}a_s)^{-1}=1$]. The intensity is normalized by $\varepsilon_{\rm F}^{-1}$. (c) Superfluid density of states $\rho(\omega)$ in Eq.~(\ref{eq.42}). $\rho_0$ is given in Eq.~(\ref{eq.xxx1}). In calculating $A({\bm k},\omega)$, we have approximated the $\delta$-function in Eq.~(\ref{eq.41}) as $\delta(x)\simeq (1/\pi)\eta/[x^2+\eta^2]$ with $\eta/\varepsilon_{\rm F}=10^{-2}$. We also use the same prescription in Figs.~\ref{fig6} and \ref{fig10}.}
\label{fig5}
\end{figure}
\par
Here, we confirm that the present odd-frequency superfluid state satisfies the positivity of the single-particle spectral weight $A({\bm k},\omega)$. Explicitly calculating $A({\bm k},\omega)$ from the $(1,1)$ component of the single-particle thermal Green's function ${\hat G}_{\rm odd}({\bm k},i\omega_n)$ in Eq.~(\ref{eq.21}), we have
\begin{align}
A({\bm k},\omega) 
&=
-{1 \over \pi}{\rm Im}
\left[
G_{\rm odd}^{(1,1)}({\bm k},i\omega_n\to\omega+i\delta)
\right]
\nonumber
\\
&=
{\Lambda^2 \over |\Delta |^2+\Lambda^2}
\delta(\omega-\xi_{\bm k})
+
{|\Delta|^2 \over |\Delta |^2+\Lambda^2}
\nonumber
\\
&\times
\left(
{1 \over 2}
\left[
1+{\xi_{\bm k} \over {\cal E}_{\rm odd}({\bm k},\Lambda)}
\right]
\delta\left[\omega-{\cal E}_{\rm odd}({\bm k},\Lambda)\right]
\right.
\nonumber
\\
&+
\left.
{1 \over 2}
\left[
1-{\xi_{\bm k} \over {\cal E}_{\rm odd}({\bm k},\Lambda)}
\right]
\delta\left[\omega+{\cal E}_{\rm odd}({\bm k},\Lambda)\right]
\right).
\label{eq.41}
\end{align}
As shown in Figs.~\ref{fig5}(a) and (b), Eq.~(\ref{eq.41}) is always positive.
\par
Equation~(\ref{eq.41}) shows that, when $\mu>0$, single-particle excitations are gapless, because of the dispersion $\omega=\xi_{\bm k}={\bm k}^2/(2m)-\mu$. Indeed, one clearly sees in Fig.~\ref{fig5}(a) that this dispersion passes through $\omega=0$. Thus, the superfluid density of states $\rho(\omega)$, which is related to $A({\bm k},\omega)$ as
\begin{equation}
\rho(\omega)=\sum_{\bm k}A({\bm k},\omega),
\label{eq.42}
\end{equation}
also becomes gapless, as shown in Fig.~\ref{fig5}(c). 
\par
In the strong-coupling regime, since the Fermi chemical potential $\mu$ becomes negative [see Figs.~\ref{fig4}(b) and (d)], the dispersion $\omega=\xi_{\bm k}={\bm k}^2/(2m)+|\mu|$ no longer passes through $\omega=0$. Because of this, when $(k_{\rm F}a_s)^{-1}=1$, gapped single-particle excitations are obtained, as seen in Figs.~\ref{fig5}(b) and (c).
\par
Regarding the above-mentioned gapless single-particle excitations, we note that the recent experimental proposal about the realization of the bulk odd-frequency superconducting state in CeRh$_{0.5}$Ir$_{0.5}$In$_5$~\cite{Kawasaki2020} is based on the observation of Korringa-law like temperature dependence of the spin-lattice relaxation rate $T_1^{-1}$ below $T_{\rm c}$. Since the Korringa law in the normal state is well-known to originate from the presence of gapless single-particle excitations around the Fermi surface, the observed anomaly implies the absence of single-particle excitation gap in this superconducting state. Thus, although our model is not directly related to this material, it is an interesting future problem to examine to what extent this simple model can explain the observed temperature dependence of $T_1^{-1}$ in CeRh$_{0.5}$Ir$_{0.5}$In$_5$.
\par
\section{Space-time structure of the odd-frequency pair wavefunction}\
\par
\begin{figure}[t]
\centering
\includegraphics[width=0.45\textwidth]{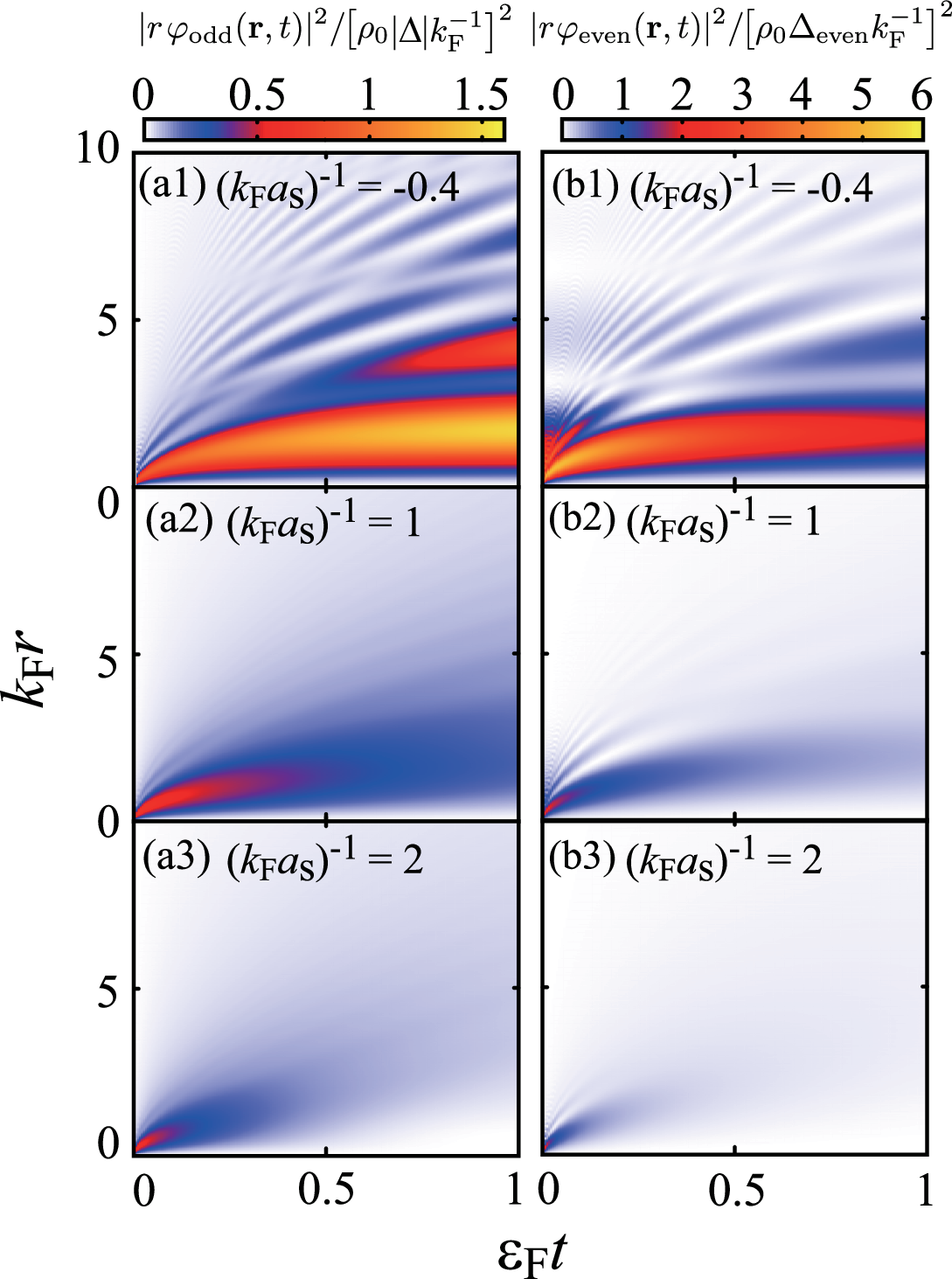}
\caption{Calculated space-time structure of the pair wavefunction at $T=0$. The left and right panels show the odd-frequency case [$\varphi_{\rm odd}({\bm r},t)$ in Eq.~(\ref{eq.x1})] and the even-frequency case [$\varphi_{\rm even}({\bm r},t)$ in Eq.~(\ref{eq.y2})], respectively. We set $\Lambda/\varepsilon_{\rm F}=0.5$.}
\label{fig6}
\end{figure}
\par
\begin{figure}[t]
\centering
\includegraphics[width=0.45\textwidth]{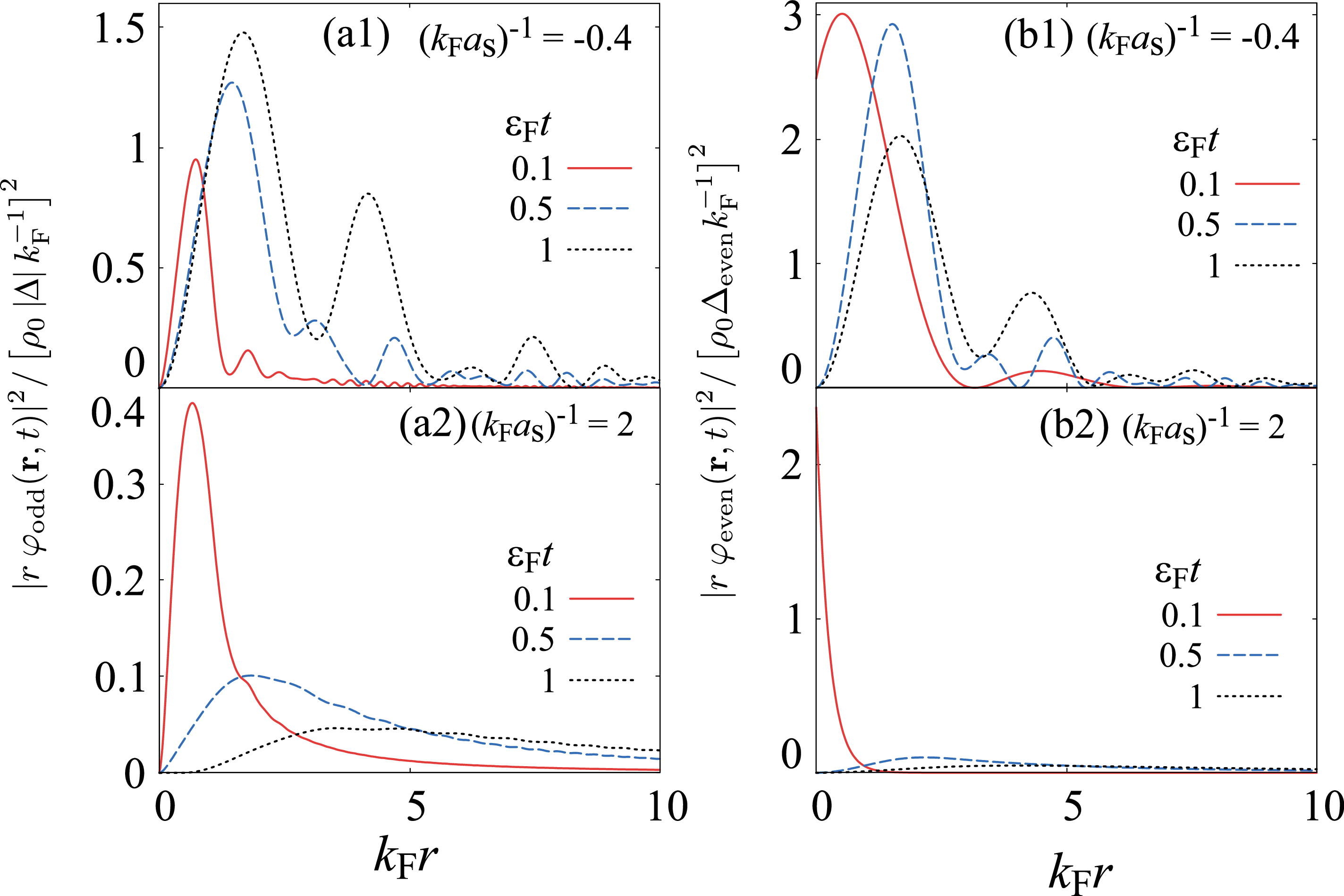}
\caption{Spatial variation of $|r\varphi_{\rm odd}({\bm r},t)|^2$ (left panels) and $|r\varphi_{\rm even}({\bm r},t)|^2$ (right panels) at $T=0$. The upper (lower) two panels show the weak-coupling case (strong-coupling case) where $\mu>0$ ($\mu<0$). We set $\Lambda/\varepsilon_{\rm F}=0.5$.}
\label{fig7}
\end{figure}
\par
Figure~\ref{fig6} shows the space-time structure of the pair wavefunction. We also show in Fig.~\ref{fig7} the detailed spatial variation of this quantity at some values of the relative time $t$ between two fermions involved in a Cooper pair.
\par
When the pairing interaction is weak [$(k_{\rm F}a_s)^{-1}=-0.4<0$], Fig.~\ref{fig6}(a1) shows that the odd-frequency pair wavefunction $\varphi_{\rm odd}({\bm r},t)$ spreads out in the temporal direction. We also find from Fig.~\ref{fig7}(a1) that $|r\varphi_{\rm odd}({\bm r},t)|^2$ has larger intensity for larger value of the relative time $t$ [at least within the temporal range shown in Fig.~\ref{fig7}(a1)]. This tendency is consistent with the fact that $\varphi_{\rm odd}({\bm r},t)$ vanishes at $t=0$  [see Eq.~(\ref{eq.x9b})]. 
\par
As shown in Fig.~\ref{fig6}(b1), the pair wavefunction $\varphi_{\rm even}({\bm r},t)$ in the even-frequency superfluid state also spreads out in the temporal direction when $(k_{\rm F}a_s)^{-1}=-0.4$; however, in contrast to the odd-frequency case, $|r\varphi_{\rm even}({\bm r},t)|^2$ has large intensity around  $t=0$, as shown in Fig.~\ref{fig7}(b1). 
\par
As the interaction strength increases, we see in Fig.~\ref{fig6} that, in both the even- and odd-frequency cases, the pair wavefunction shrinks to gather around the origin of the space-time plane ($r=t=0$). In the strong-coupling regime when $(k_{\rm F}a_s)^{-1}=2$, in spite of $\varphi_{\rm odd}({\bm r},t=0)=0$, $|r\varphi_{\rm odd}({\bm r},t)|^2$ has large intensity around $k_{\rm F}r=1$ at the relatively short relative time $\varepsilon_{\rm F}t=0.1~(\ll 1)$, as shown in Fig.~\ref{fig7}(a2). This tendency is more remarkable in the even-frequency case shown in Fig.~\ref{fig7}(b2), where $|r\varphi_{\rm even}({\bm r},t)|^2$ is almost dominated by the intensity at $t=0$. As mentioned previously, because the even-frequency pair wavefunction at $t=0$ is reduced to the wavefunction of a two-body bound state when the pairing interaction is very strong, this result indicates that the pair wavefunction $\varphi_{\rm even}({\bm r},t)$ may almost be viewed as the wavefunction of a two-body bound state given in Eq.~(\ref{eq.x6}), when $(k_{\rm F}a_s)^{-1}=2$.
\par
Here, we comment on the spatial oscillation of the pair wavefunction seen in the upper panels in Fig.~\ref{fig7}: As found from the factor $\sin(k_{\rm F}r)$ involved in Eq.~(\ref{eq.x7}), this oscillation originates from the existence of the Fermi surface. Thus, with increasing the interaction strength, such oscillating behavior of the pair wavefunction becomes obscure due to the decrease of the Fermi chemical potential [see Fig.~\ref{fig4}(b)], which may be interpreted as the shrinkage of the Fermi surface size. Then, since the negative chemical potential realized in the strong-coupling regime can be interpreted as the disappearance of the Fermi surface, the oscillation of the pair wavefunction also disappears in this regime, as seen in the lower panels in Fig.~\ref{fig7}.
\par
\begin{figure}[t]
\centering
\includegraphics[width=0.45\textwidth]{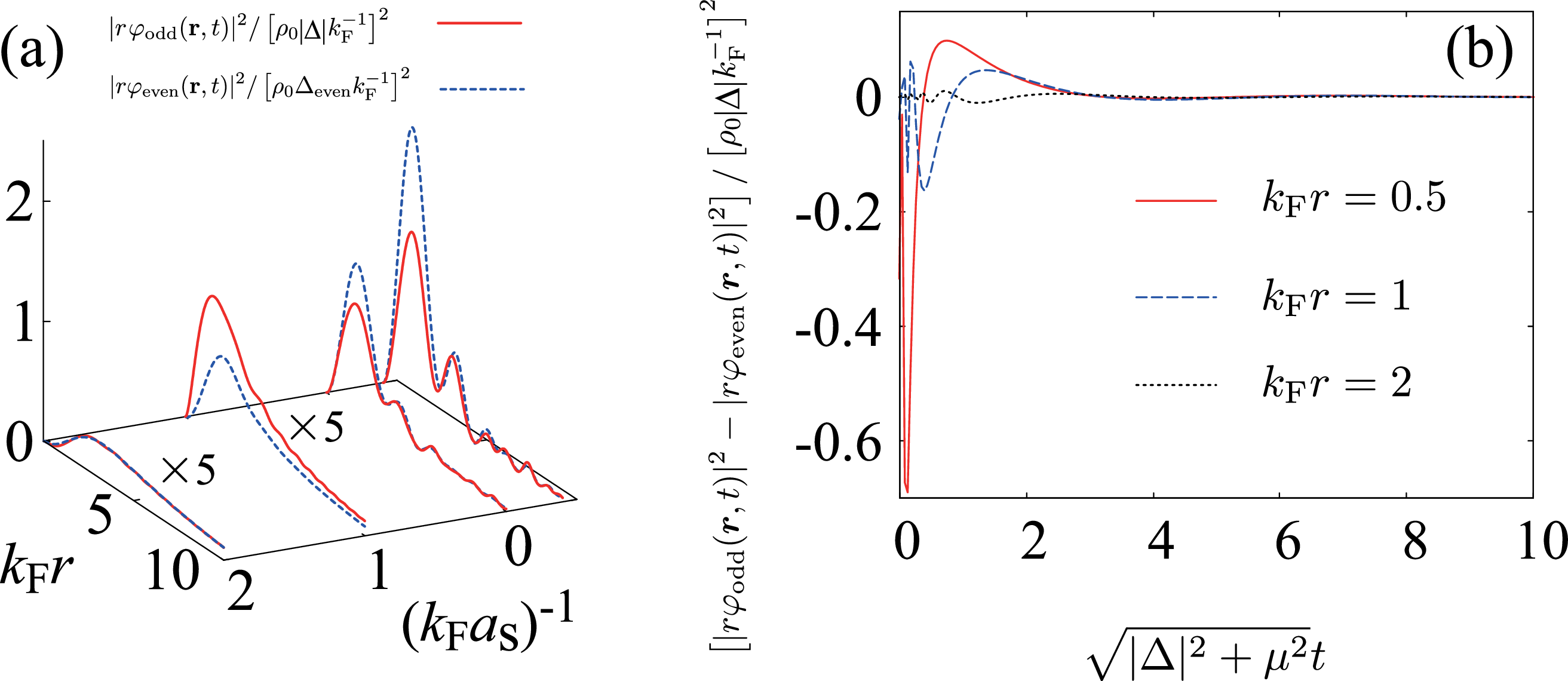}
\caption{Comparison of $|r\varphi_{\rm odd}({\bm r},t)|^2$ with $|r\varphi_{\rm even}({\bm r},t)|^2$ at $T=0$. (a) We take $\Lambda/\varepsilon_{\rm F}=0.5$ and $\varepsilon_{\rm F}t=0.8$. When $(k_{\rm F}a_s)^{-1}=1$ and 2, results are magnified by the factor 5. (b) We take $\Lambda/\varepsilon_{\rm F}=0.5$ and $(k_{\rm F}a_s)^{-1}=2$.}
\label{fig8}
\end{figure}

Figure~\ref{fig8} compares the magnitude of the pair wavefunction in the odd-frequency superfluid state with that in the even-frequency case. At $\varepsilon_{\rm F}t=0.8$ shown in Fig. \ref{fig8}(a), when the pairing interaction is relatively weak [$(k_{\rm F}a_s)^{-1}\le 0$], the difference between the two is remarkable, especially around $k_{\rm F}r=1$. Their difference becomes small as the interaction strength increases, as seen in Fig.~\ref{fig8}(a). 
\par
We also show in Fig. \ref{fig8}(b) the temporal dependence of the difference between these quantities in the strong-coupling regime [$(k_{\rm F}a_s)^{-1}=2$]. As analytically shown in Appendix D, these quantities become close to each other when $t\gg1/\sqrt{|\Delta|^2+\mu^2}$, which can be confirmed in this figure. In the strong coupling limit $\left[(k_{\rm F}a_s)^{-1}\to\infty\right]$, since $|\Delta|$ and $|\mu|$ diverge, this condition is always satisfied except at $t=0$. Thus, in this limit, one finds
\begin{equation}
\varphi_{\rm odd}({\bm r},t)=-{\rm sgn}(t)\varphi_{\rm even}({\bm r},t).~~~~(t\ne0)
\label{eq.43}
\end{equation}
(For the derivation, see Appendix D. ) That is, the {\it magnitude} $|\varphi_{\rm odd}({\bm r},t)|$ of the odd-frequency pair wavefunction in the strong-coupling limit has the same space-time structure as the even-frequency case, except at $t=0$.
\par
\par
\begin{figure}[t]
\centering
\includegraphics[width=0.45\textwidth]{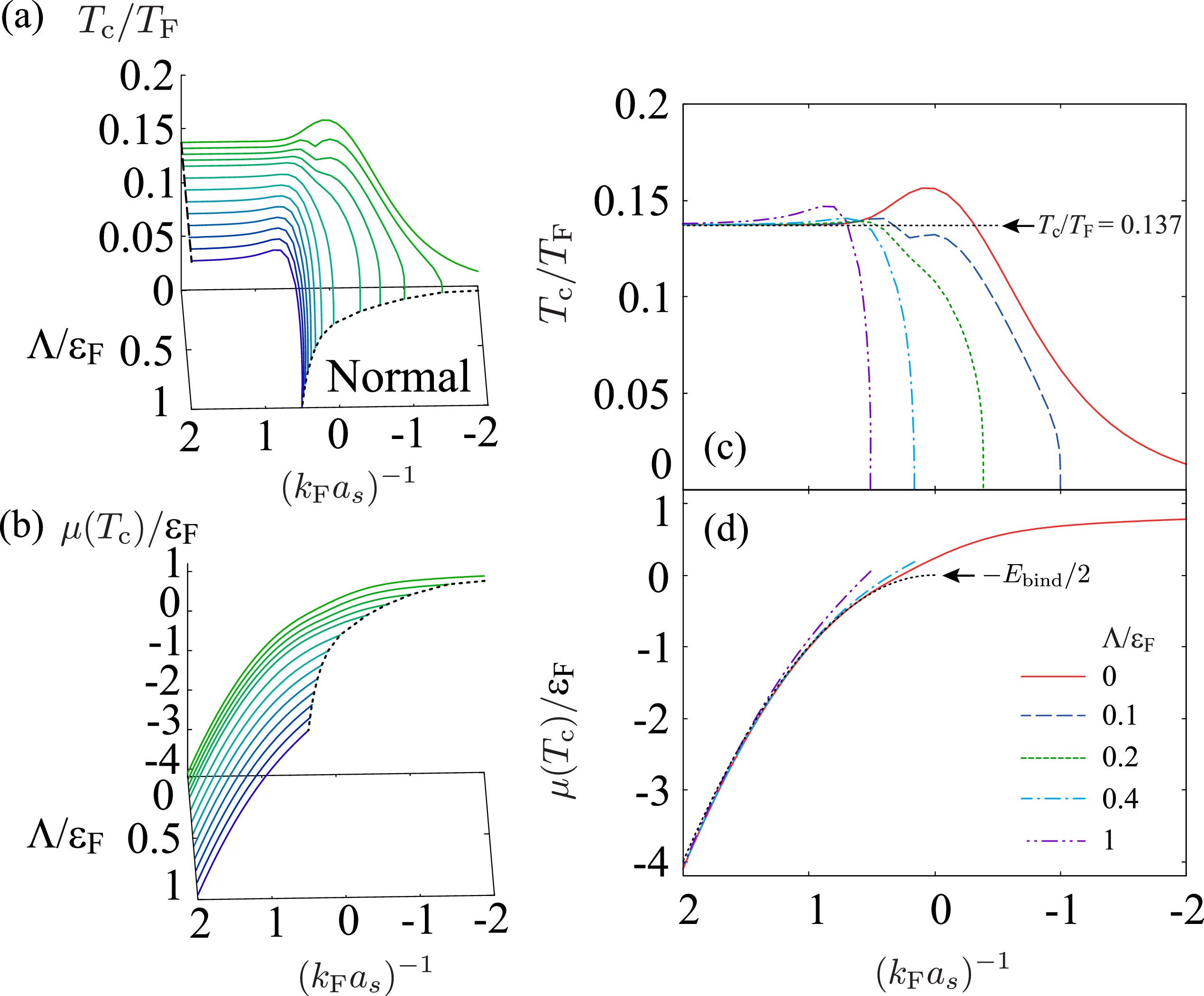}
\caption{Calculated (a) $T_{\rm c}$ and (b) $\mu(T_{\rm c})$ as functions of $\Lambda$ and the interaction strength $(k_{\rm F}a_s)^{-1}$, in a single-component Fermi gas with an odd-frequency pairing interaction. The dashed line is the BEC phase transition temperature $T_{\rm BEC}=0.137T_{\rm F}$. The dotted line shows the threshold interaction strength $(k_{\rm F}a_s^{\rm c})^{-1}$ given in Eq.~(\ref{eq.46}). Panels (c) and (d), respectively, show detailed interaction dependence of $T_{\rm c}$ and $\mu(T_{\rm c})$. In panel (d), 
since the results for $\Lambda/\varepsilon_{\rm F}=0.1$ and 0.2 are almost the same as that for $\Lambda/\varepsilon_{\rm F}=0.4$, we only show the results for $\Lambda/\varepsilon_{\rm F}=0, 0.4$, and 1. }
\label{fig9}
\end{figure}

Since the even-frequency pair wavefunction in the strong-coupling regime may be viewed as the wavefunction of a two-body bound state, Fig.~\ref{fig8}, as well as Eq.~(\ref{eq.43}), make us expect that, as in the even-frequency case, the superfluid phase transition into the odd-frequency superfluid state may also be described by the BEC of molecular bosons in the strong-coupling regime, unless the absence of the equal-time pairing and the sign change of $\varphi_{\rm odd}({\bm r},t)$ at $t=0$ seriously affect the superfluid instability. To confirm this expectation, we show in Figs.~\ref{fig9}(a) and (c) the calculated $T_{\rm c}$  within the framework of the NSR theory explained in Sec. II.D. We see in these figures that, with increasing the interaction strength, $T_{\rm c}$ always approaches the expected BEC phase transition temperature $T_{\rm BEC}$ of an ideal Bose gas with the molecular number $N_{\rm B}=N/2$ and the molecular mass $M_{\rm B}=2m$:
\begin{equation}
T_{\rm BEC}=
{2\pi \over 2m}
\left({N \over 2\zeta(3/2)}\right)^{2/3}=0.137T_{\rm F},
\label{eq.44}
\end{equation}
where $\zeta(3/2)=2.612$ is the zeta function. This clearly indicates that, as in the even-frequency case, odd-frequency Cooper pairs also behave like `bosons' in this regime, in spite of the absence of equal-time Cooper pairing.
\par
\par
Similarity between the even- and odd-frequency cases can also be seen in Figs.~\ref{fig9}(b) and (d): In the strong-coupling regime, the Fermi chemical potential $\mu$ becomes negative and the magnitude $|\mu|$ approaches half the binding energy $E_{\rm bind}$ of a two-body bound state given in Eq.~(\ref{eq.xx6}). The Fermi chemical potential physically means the energy to add a particle to the system. One finds from this behavior of $\mu$ that, as in the even-frequency case, an odd-frequency Cooper pair in the strong-coupling regime also has the binding energy $E_{\rm bind}$ given in Eq.~(\ref{eq.xx6}). 
\par
We point out that the above-mentioned bosonic character comes from the structure of the NSR number equation~(\ref{eq.34}): Since $|\mu|\gg\Lambda$ in the strong-coupling regime, the basis function in Eq.~(\ref{eq.6}) can be approximated as $\gamma({\bm k},i\omega_n)\simeq {\rm sgn}(\omega_n)$ there. Then the pair-correlation function in Eq.~(\ref{eq.32}), as well as the resulting NSR number equation~(\ref{eq.34}), have the same forms as those in the ordinary (even-frequency) $s$-wave superfluid Fermi gas discussed in BCS-BEC physics~\cite{NSR1985,Melo1993,Randeria1995,Ohashi2002,Strinati2002,Chen2005,Giorgini2008,Bloch2008,Ohashi2020}. Thus, as is well known in the standard NSR theory~\cite{NSR1985}, the number equation~(\ref{eq.34}) in the strong-coupling regime is reduced to
\begin{equation}
{N \over 2}=\sum_{\bm q}
n_{\rm B}
\left(
{{\bm q}^2 \over 2M_{\rm B}}-\mu_{\rm B}
\right),
\label{eq.45}
\end{equation}
where $n_{\rm B}(x)$ is the Bose distribution function, and $\mu_{\rm B}=2\mu+E_{\rm bind}$ plays the role of the Bose chemical potential. Equation~(\ref{eq.45}) immediately gives $T_{\rm BEC}$ in Eq.~(\ref{eq.44}), when $\mu=-E_{\rm bind}/2$, being consistent with the strong-coupling behavior of $T_{\rm c}$ and $\mu(T_{\rm c})$ shown in Fig.~\ref{fig9}~\cite{note3}.
\par
To conclude, odd-frequency Cooper pairs also behave like molecular bosons in the strong-coupling regime. Although this result is the same as the case of the ordinary (even-frequency) $s$-wave superfluid Fermi gas in the BCS-BEC crossover region~\cite{NSR1985,Melo1993,Randeria1995,Ohashi2002,Strinati2002,Chen2005,Giorgini2008,Bloch2008,Ohashi2020}, the odd-frequency pair wavefunction $\varphi_{\rm odd}({\bm r},t)$ itself is {\it not} completely the same as the even-frequency pair wavefunction $\varphi_{\rm even}({\bm r},t)$ in this regime, as shown in Eq.~(\ref{eq.43}). This means that the sign change of the odd-frequency pair wavefunction, as well as the absence of equal-time pairing (that are different from the even-frequency case), are not crucial for the odd-frequency Cooper pair to possess bosonic character, at least in considering the superfluid phase transition temperature.
\par
Before ending this section, we discuss the behavior of $T_{\rm c}$ in the weak-coupling regime: When $\Lambda=0$, which gives $\gamma({\bm k},i\omega_n)={\rm sgn}(\omega_n)$, the $T_{\rm c}$ equation~(\ref{eq.35}) and the number equation~(\ref{eq.34}) coincide with those in the ordinary NSR theory for the even-frequency $s$-wave superfluid Fermi gas. Thus, the system always experiences the superfluid instability at $T_{\rm c}>0$, irrespective of the value of $(k_{\rm F}a_s)^{-1}$ [see Figs.~\ref{fig9}(a) and (c)]. On the other hand, when $\Lambda>0$, as expected from the zero-temperature result shown in Fig.~\ref{fig4}, $T_{\rm c}$ vanishes, when the interaction strength becomes weaker than the threshold value,
\begin{equation}
(k_{\rm F}a_s^{\rm c})^{-1}
={2 \over \pi}\ln
\left(
{e^2 \over 8}{\Lambda \over \mu}
\right),
\label{eq.46}
\end{equation}
which is obtained by simply setting $T_{\rm c}=0$ in the $T_{\rm c}$ equation~(\ref{eq.35})~\cite{note4}. 
\par
To quickly grasp how $\Lambda$ affects $T_{\rm c}$, it is convenient to approximately deal with the gap equation~(\ref{eq.27}) in the weak-coupling regime at $T_{\rm c}$ as
\begin{eqnarray}
1
&=&
UT_{\rm c}\sum_{{\bm k},\omega_n}
{1 \over \omega_n(T_{\rm c})^2+\xi_{\bm k}^2+\Lambda^2}
\nonumber
\\
&\simeq&
U\rho_0T_{\rm c}
\sum_{\omega_n}\int_{-\infty}^\infty d\xi
{1 \over \omega_n(T_{\rm c})^2+\xi^2+\Lambda^2}
\nonumber
\\
&=&
\pi U\rho_0T_{\rm c}\sum_{\omega_n}
{1 \over \sqrt{\omega_n(T_{\rm c})^2+\Lambda^2}},
\label{eq.46b}
\end{eqnarray}
where the density of states $\rho_0$ is given in Eq.~(\ref{eq.xxx1}). Writing the superfluid phase transition temperature at $\Lambda=0$ as $T_{{\rm c}0}$, we rewrite the $T_{\rm c}$ equation~(\ref{eq.46b}) as
\begin{equation}
\ln
\left(
{T_{\rm c} \over T_{{\rm c}0}}
\right)
=T_{\rm c}
\sum_{\omega_n}
\left[
{1 \over \sqrt{\omega_n(T_{\rm c})^2+\Lambda^2}}
-
{1 \over |\omega_n(T_{\rm c})|}
\right].
\label{eq.47}
\end{equation}
Expanding the right-hand side in Eq.~(\ref{eq.47}) up to $O(\Lambda^2)$ by assuming $\Lambda\ll T_{{\rm c}0}$, one has, after summing up the Matsubara frequencies,
\begin{equation}
T_{\rm c}=T_{{\rm c}0}
\left[
1-{7\zeta(3) \over 8\pi^2}
\left(
{\Lambda \over T_{{\rm c}0}}
\right)^2
\right],
\end{equation}
where we have approximated the left-hand side in Eq.~(\ref{eq.47}) as $\ln(T_{\rm c}/T_{{\rm c}0})\simeq (T_{\rm c}/T_{{\rm c}0})-1$.
\par
\par
\section{Summary}
\par
To summarize, we have discussed pairing properties of an odd-frequency superfluid Fermi gas. This superfluid has the unique property that the Cooper pairs are formed between fermions at different times. In this paper, we examined whether or not such odd-frequency Cooper pairs still behave like molecular bosons in the strong-coupling regime, where this picture is known to be valid for the even-frequency $s$-wave superfluid system. For this purpose, we proposed a model odd-frequency pairing interaction that satisfies the positivity of the single-particle excitation spectrum. To avoid the well-known puzzle that the odd-frequency superfluid state unphysically becomes stable {\it above} $T_{\rm c}$ in the Hamiltonian formalism, we employed the recently proposed prescription using the path-integral formalism~\cite{Solenov2009,Kusunose2011B}.
\par
We calculated the space-time structure of the pair wavefunction $\varphi_{\rm odd}({\bm r},t)$ in the odd-frequency superfluid state at $T=0$, within the framework of the strong-coupling theory developed by Eagles and Leggett. From the comparison with the pair wavefunction $\varphi_{\rm even}({\bm r},t)$ in the even-frequency $s$-wave superfluid state, we found that, while $\varphi_{\rm odd}({\bm r},t)$ has different space-time structure from $\varphi_{\rm even}({\bm r},t)$ in the sense that the former always vanishes at $t=0$ and changes its sign at this time, the magnitude $|\varphi_{\rm odd}({\bm r},t)|$ becomes close to $\varphi_{\rm even}({\bm r},t)$ in the strong-coupling regime when $|t|\gg 1/\sqrt{|\Delta|^2+\mu^2}$. Particularly in the strong-coupling limit (where $|\Delta|$ and $|\mu|$ diverge), one obtains $|\varphi_{\rm odd}({\bm r},t)|=\varphi_{\rm even}({\bm r},t)$ except at $t=0$. Since the even-frequency pair wavefunction in the strong-coupling regime is dominated by the equal-time component, which is just the same as the wavefunction of a two-body bound molecule, this coincidence makes us expect that odd-frequency Cooper pairs may also have bosonic character. 
\par
To confirm this expectation, we calculated the superfluid phase transition temperature $T_{\rm c}$, by extending the NSR strong-coupling theory for the even-frequency $s$-wave Fermi superfluid to the odd-frequency case. The calculated $T_{\rm c}$ in the strong-coupling regime was found to approach the expected BEC phase transition temperature $T_{\rm BEC}$ in an ideal molecular Bose gas, which confirms that the odd-frequency Cooper pairs indeed behave like bosons there. This indicates that, although odd-frequency superfluids do not have equal-time pairing and the odd-frequency wavefunction does not coincide with the even-frequency one, the odd-frequency Cooper pair still possesses bosonic character when the condition $|\varphi_{\rm odd}({\bm r},t\ne 0)|\simeq \varphi_{\rm even}({\bm r},t\ne 0)$ is satisfied in the strong-coupling regime.
\par
The above conclusion indicates that, considering a two-component Fermi gas with the {\it even}-frequency pairing interaction given in Eq.~(\ref{eq.4}) where the basis function $\gamma({\bm k},i\omega_n)$ is replaced by $|\gamma({\bm k},i\omega_n)|$, one reproduces the same results obtained in this paper. Thus, it would be an interesting future problem to explore a phenomenon which is {\it sensitive to the sign change of} $\varphi_{\rm odd}({\bm r},t)$, in order to highlight the character of the odd-frequency pairing state.
\par
Although we have only considered the specific interaction in Eqs. (\ref{eq.4}) and (\ref{eq.6}) in this paper, our model is still expected to capture universal low-energy properties of odd-frequency Fermi superfluids where the low-frequency behavior of the superfluid order parameter behaves as $\Delta({\bm k},i\omega_n)\propto \omega_n$. For example, such linear-$\omega_n$ behavior of the odd-frequency superfluid order parameter in the low-frequency region can be realized in an electron-phonon model, where the increase of $T_{\rm c}$ with decreasing the frequency of the Einstein phonon, which just corresponds to the decrease of $\Lambda$ in our model, is predicted~\cite{Kusunose2011A}. This behavior of $T_{\rm c}$ is consistent with our results in the weak-coupling regime, shown in Fig.~9(a). In addition, it has been shown in a two-band Hubbard model that $T_{\rm c}$ in the strong-coupling regime of the odd-frequency pairing state agrees well with the ordinary BEC phase transition temperature in a molecular Bose gas~\cite{Inokuma2024}, which also agrees with our result. On the other hand, while the model interaction in Eq.~(\ref{eq.4}) gives a non-zero constant value of the superfluid order parameter in the high-frequency limit [see, Eqs. (\ref{eq.5}) and (\ref{eq.6})], the superfluid order parameter vanishes in this limit in the above-mentioned electron-phonon case~\cite{Kusunose2011A}. Thus, one needs to carefully check model dependence for superfluid properties that are sensitive to detailed high-frequency behavior of the superfluid order parameter. It remains as a future problem to clarify how detailed high-frequency structure of the pairing interaction affects physical properties of odd-frequency Fermi superfluids.
\par
In this paper, we simply assumed a model odd-frequency pairing interaction in order to examine the character of Cooper pairs in the strong-coupling regime. Thus, another crucial future problem is to explore a more fundamental model that gives the effective interaction assumed in this paper. Since various odd-frequency pairing mechanisms have recently been proposed in both metallic superconductivity~\cite{Balatsky1992,Emery1992,Abrahams1993,Asano2007,Yokoyama2007,Yokoyama2011,Tanaka2012,Cayao2020,Fuseya2003,Shigeta2009,Kusunose2011A,Matsumoto2012,Kusunose2012,Inokuma2024,Tsvelik1993,Tsvelik1994,Hoshino2014A,Hoshino2014B,Funaki2014,BlackSchaffer2013,Triola2020,Triola2016,Triola2017,Cayao2021,Miki2021,Bernardo2015A,Bernardo2015B,Kawasaki2020} and ultracold Fermi gases~\cite{Kalas2008,Arzamasovs2018,Linder2019}, it is also an interesting problem to examine how the present model is related to these proposals. We also note that, although we have presented an example of the interaction that gives the required positive single-particle spectral weight, clarifying the general condition for the odd-frequency pairing interaction to satisfy this requirement still remains to be solved.  
\par
For the superfluid state, we have only examined the cases at $T_{\rm c}$ and the $T=0$ in this paper. Regarding this, we note that the Korringa-law-like temperature dependence of the spin-lattice relaxation rate $T_1^{-1}$ has recently been observed in heavy fermion superconductor CeRh$_{0.5}$Ir$_{0.5}$In$_5$~\cite{Kawasaki2020}. Based on this observation, Ref.~\cite{Kawasaki2020} proposed the realization of the odd-frequency superconducting state with a gapless superconducting density of states in this material. (Note that the Korringa law in the normal state originates from the existence of gapless single-particle excitations around the Fermi level.) Since the odd-frequency superfluid state discussed in this paper gives gapless single-particle excitations (see Fig.~\ref{fig5}), the extension of our theory to the superfluid phase below $T_{\rm c}$ would enable us to examine to what extent the observed $T$-linear behavior of $T_1^{-1}$ can be explained in our model. Since odd-frequency superfluids have recently been discussed in both condensed matter physics and cold atom physics, our results would contribute to the further development of these active research fields. 
\par
\par
\begin{acknowledgments}
We thank H. Kusunose, D. Kagamihara, and D. Inotani for discussions. S.I. was supported by a Grant-in-Aid for JST SPRING (Grant No.JPMJSP2123). T.K. was supported by MEXT and JSPS KAKENHI Grantin-Aid for JSPS fellows (Grant No. JP21J22452). Y.O. was supported by a Grant-in-Aid for Scientific Research from MEXT and JSPS in Japan (Grant No.JP18K11345, No.JP18H05406, No.JP19K03689, and JP22K03486).
\end{acknowledgments}
\par
\appendix
\par
\section{Separable odd-frequency pairing interaction}
\par
Here, we present an example that gives a separable pairing interaction. When a Fermi-Fermi interaction is mediated by the Einstein phonon with the frequency $\Omega_{\rm E}$, the interaction $V_{k,k'}$ appearing in the action $S_1$ in Eq.~(\ref{eq.3}) is given by~\cite{Kusunose2011A}
\begin{equation}
V_{k,k'}=g^2D(k-k'),
\label{eq.A1}
\end{equation}
where $g$ is a fermion-phonon coupling constant, and 
\begin{equation}
D(q)=-{\Omega_{\rm E}^2 \over \nu_n^2+\Omega_{\rm E}^2}
\label{eq.A2}
\end{equation}
is the free phonon Green's function~\cite{Mahan}. When we write Eq.~(\ref{eq.A1}) as
\begin{eqnarray}
V(k,k')
&=&{g^2 \over 2}[D(k-k')+D(k+k')]
\nonumber
\\
&&+
{g^2 \over 2}[D(k-k')-D(k+k')]
\nonumber
\\
&\equiv& V_+(k,k')+V_-(k,k'),
\label{eq.A3}
\end{eqnarray}
the odd-frequency Cooper channel is given by the latter part $V_-(k,k')=(g^2/2)[D(k-k')-D(k+k')]$~\cite{Kusunose2011A}. Only retaining this, one reaches
\begin{equation}
V(k,k')=-{2g^2 \over \Omega_{\rm E}^2}
{\omega_n\omega_n' 
\over
\displaystyle
\left[1+\left({\omega_n-\omega_n' \over \Omega_{\rm E}}\right)^2\right]
\left[1+\left({\omega_n+\omega_n' \over \Omega_{\rm E}}\right)^2\right]
}.
\label{eq.A4}
\end{equation}
Expanding the denominator in Eq.~(\ref{eq.A4}) up to $O((\omega_n/\Omega_{\rm E})^2)$ and $O((\omega_n'/\Omega_{\rm E})^2)$, one can rewrite Eq.~(\ref{eq.A4}) into the separable form as
\begin{eqnarray}
V(k,k')
&\simeq& -{2g^2 \over \Omega_{\rm E}^2}
{\omega_n\omega_n' 
\over
1+
2(\omega_n/\Omega_{\rm E})^2+
2(\omega_n'/\Omega_{\rm E})^2
}
\nonumber
\\
&\simeq&
-{g^2 \over 2}
{
\omega_n 
\over 
\sqrt{\omega_n^2+(\Omega_{\rm E}/2)^2}
}
{
\omega_n' 
\over 
\sqrt{\omega_n'^2+(\Omega_{\rm E}/2)^2}
}
\nonumber
\\
&\equiv&
-{g^2 \over 2}{\tilde \gamma}(i\omega_n){\tilde \gamma}(i\omega_n').
\label{eq.A5}
\end{eqnarray}
\par
We briefly note that, setting $\Lambda=\Omega_\mathrm{E}/2$ and $\xi_{\bm k}=0$ in Eq.~(\ref{eq.6}), we find that $\tilde{\gamma}(i\omega_n)=\gamma({\bm k},i\omega_n)$. This implies that $\Lambda$ plays a similar role to the frequency of the Einstein phonon.
\par
\begin{figure}[t]
\centering
\includegraphics[width=0.4\textwidth]{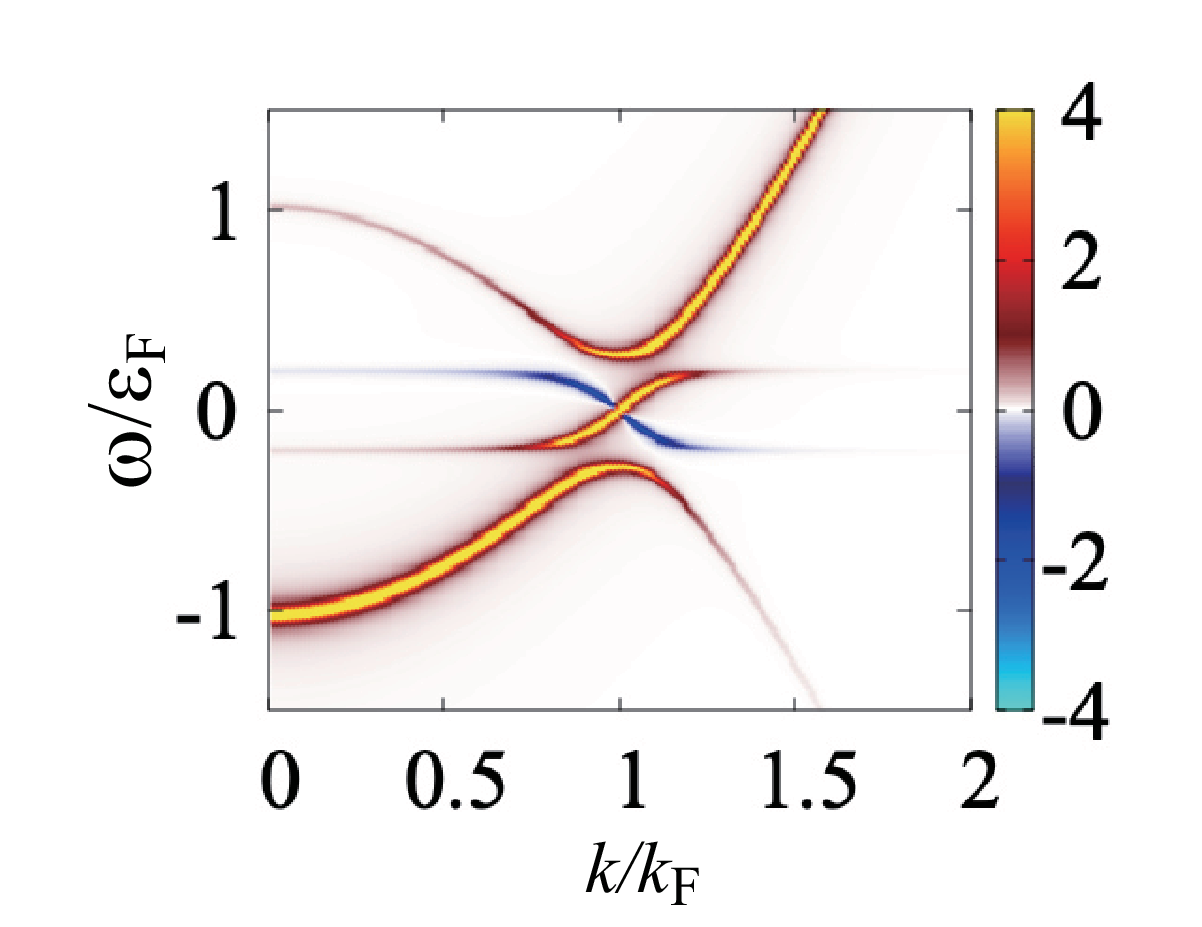}
\caption{Calculated single-particle spectral weight $A({\bm k},\omega)$ in Eq.~(\ref{eq.A6}). We set $\mu/\varepsilon_{\rm F}=1$, $\Lambda/\varepsilon_{\rm F}=0.1$, and $|\Delta|/\varepsilon_{\rm F}=0.2$.}
\label{fig10}
\end{figure}
\par
\par
The basis function ${\tilde \gamma}(i\omega_n)$ is essentially the same as Eq.~(\ref{eq.4}) where $\xi_{\bm k}$ is ignored. Regarding this, we note that this simpler version of the basis function, however, does not satisfy the positivity of the single-particle spectral weight. Indeed, when we ignore $\xi_{\bm k}$ in Eq.~(\ref{eq.4}), the resulting spectral weight,  
\begin{equation}
A({\bm k},\omega)=
-{1 \over \pi}
{\rm Im}
\left[
{\omega+i\delta+\xi_{\bm k}
\over
(\omega+i\delta)^2-\xi_{\bm k}^2-
|\Delta|^2
{(\omega+i\delta)^2 \over (\omega+i\delta)^2-\Lambda^2}
}
\right],
\label{eq.A6}
\end{equation}
has a negative branch around $\omega=0$, as shown in Fig.~\ref{fig10}.
\par
\par
\section{Stratonovich-Hubbard transformation}
\par
We explain how to derive Eqs.~(\ref{eq.18}) and (\ref{eq.19}).
We introduce the Cooper-pair Bose field $\Phi$ as well as its conjugate field ${\bar \Phi}$, by way of the Stratonovich-Hubbard transformation~\cite{Stratonovich1958,Hubbard1959}. The partition function in Eq.~(\ref{eq.1}) is then transformed as
\begin{equation}
Z\propto
\int{\cal D}{\bar \Phi}{\cal D}\Phi
\int{\cal D}{\bar \psi}{\cal D}\psi 
e^{-S[{\bar \psi},\psi,{\bar \Phi},\Phi]},
\label{eq.9}
\end{equation}
where ${\cal D}{\bar \Phi}{\cal D}\Phi$ is given in Eq.~(\ref{eq.18}), and the action $S[{\bar \psi},\psi,{\bar \Phi},\Phi]$ has the form,
\begin{equation}
S[{\bar \psi},\psi,{\bar \Phi},\Phi]=
S_0-
\frac{1}{2}\sum_q
\left[
{{\bar \Phi}_q \Phi_q \over U}
-{\bar \rho}_q \Phi_q
-\rho _q{\bar \Phi}_q
\right],
\label{eq.11}
\end{equation}
\begin{equation}
\rho_q=
\sum_k \frac{1}{\sqrt{\beta}}
\gamma\left({\bm k}+\frac{{\bm q}}{2},i\omega_n + \frac{i\nu_m}{2} \right)
\psi_{-k}
\psi_{k+q},
\label{eq.12}
\end{equation}
\begin{equation}
{\bar \rho}_q=\sum_k
\sum_k \frac{1}{\sqrt{\beta}}
\gamma\left({\bm k}+\frac{{\bm q}}{2},i\omega_n + \frac{i\nu_m}{2} \right)
{\bar \psi}_{k+q}
{\bar \psi}_{-k}.
\label{eq.13}
\end{equation}
Introducing the two-component Nambu fields,
\begin{eqnarray}
{\hat \Psi}_k\equiv
\left(
\begin{array}{c}
\psi_k \\
{\bar \psi}_{-k}
\end{array}
\right),
\label{eq.14}
\end{eqnarray}
\begin{equation}
{\hat \Psi}_{k}^\dagger
\equiv
\left({\bar \psi}_k,\psi_{-k}\right),
\label{eq.15}
\end{equation}
we rewrite Eq.~(\ref{eq.11}) as~\cite{Randeria2008,Stoof2009},
\begin{equation}
S=
{1 \over 2}\sum_k {\hat \Psi}_k^\dagger
\left[-{\hat G}^{-1}_{kk'}\right]
{\hat \Psi}_{k'}
+
\sum_q {{\bar \Phi}_q \Phi_q \over 2U}
+
{\beta \over 2}\sum _{\bm k}\xi_{\bm k}.
\label{eq.16}
\end{equation}
Here, ${\hat G}^{-1}_{kk'}$ is the inverse of the $2\times 2$ matrix single-particle thermal Green's function given in Eq.~(\ref{eq.17}). Carrying out the fermion path integrals in Eq.~(\ref{eq.9}), one obtains Eqs. (\ref{eq.18}), and (\ref{eq.19}).
\par
\par
\section{Even-frequency pair wavefunction}
\par
We derive Eqs.~(\ref{eq.y1}) and (\ref{eq.y2}). We assume a two-component Fermi gas described by the standard BCS Hamiltonian:
\begin{align}
H_{\rm BCS}&=
\int d{\bm r}
\sum_{\sigma=\uparrow,\downarrow}
\psi^\dagger_\sigma({\bm r})
\left[
-{{\bm \nabla}^2 \over 2m}-\mu_{\rm even}
\right]
\psi_\sigma({\bm r})
\nonumber
\\
&-U\int d{\bm r}
\psi_\uparrow^\dagger({\bm r})
\psi_\downarrow^\dagger({\bm r})
\psi_\downarrow({\bm r})
\psi_\uparrow({\bm r})
,
\end{align}
where the field operator $\psi_{\sigma=\uparrow,\downarrow}({\bm r})$ describes fermions with pseudospin $\sigma=\uparrow,\downarrow$. The corresponding $2\times 2$ matrix mean-field BCS single-particle thermal Green's function is given by
\begin{equation}
{\hat G}_{\rm even}({\bm k},i\omega_n)=
{1 \over i\omega_n-\xi_{\rm even}({\bm k})\tau_3+\Delta_{\rm even}\tau_1}.
\label{eq.x4}
\end{equation}
Here, $\xi_{\rm even}({\bm k})=\varepsilon_{\bm k}-\mu_{\rm even}$ is the kinetic energy, measured from the Fermi chemical potential $\mu_{\rm even}$. The $s$-wave superfluid order parameter $\Delta_{\rm even}$ (which is taken to be real, for simplicity), as well as $\mu_{\rm even}$, are determined from the BCS-Eagles-Leggett coupled equations given in Eqs.~(\ref{eq.x4b}) and (\ref{eq.x4c}).
\par
The equal-time component $\varphi_{\rm even}({\bm r})$ of the even-frequency wavefunction, which is given in the first line in Eq.~(\ref{eq.y1}), is related to the lesser Green's function ${\cal G}_{\rm even}^{<,(1,2)}({\bm k},\omega)$ as 
\begin{eqnarray}
\varphi_{\rm even}({\bm r})
&=&
-i\sum_{\bm k}\int_{-\infty}^\infty{d\omega \over 2\pi}
e^{-i{\bm k}\cdot{\bm r}}
{\cal G}_{\rm even}^{<,(1,2)}({\bm k},\omega).
\label{eq.x3}
\end{eqnarray}
Here, ${\cal G}_{\rm even}^{<,(1,2)}({\bm k},\omega)$ is obtained from ${\hat G}_{\rm even}({\bm k},i\omega_n)$ in Eq.~(\ref{eq.x4}) as 
\begin{eqnarray}
{\cal G}_{\rm even}^{<,(1,2)}({\bm k},\omega)
&=&-f(\omega)
\left[
G_{\rm even}^{(1,2)}({\bm k},i\omega_n\to\omega+i\delta)|_{\omega_n>0}
\right.
\nonumber
\\
&&-
\left.
G_{\rm even}^{(1,2)}({\bm k},i\omega_n\to\omega-i\delta)|_{\omega_n<0}
\right].
\label{eq.x5}
\end{eqnarray}
Substituting Eqs.~(\ref{eq.x4}) and~(\ref{eq.x5}) into Eq.~(\ref{eq.x3}), and executing the $\omega$ integration in Eq.~(\ref{eq.x3}), one reaches the second line in Eq.~(\ref{eq.y1}) at $T=0$. 
\par
The time-dependent even-frequency pair wavefunction given in the first line in Eq.~(\ref{eq.y2}) is related to the lesser Green's function as 
\begin{eqnarray}
\varphi_{\rm even}({\bm r},t)
&=&
-i\sum_{\bm k}e^{-i{\bm k}\cdot{\bm r}}
{\cal G}_{\rm even}^{<,(1,2)}({\bm k},-t),
\label{eq.x10}
\end{eqnarray}
where
\begin{equation}
{\cal G}_{\rm even}^{<,(1,2)}({\bm k},t)=
\int_{-\infty}^\infty{d\omega \over 2\pi}
e^{-i\omega t}
{\cal G}_{\rm even}^{<,(1,2)}({\bm k},\omega).
\label{eq.x10b}
\end{equation}
Substituting Eqs.~(\ref{eq.x4}) and~(\ref{eq.x5}) into Eq.~(\ref{eq.x10b}), and carrying out the $\omega$ integration in Eq.~(\ref{eq.x10b}), we obtain, at $T=0$, 
\begin{equation}
{\cal G}_{\rm even}^{<,(1,2)}({\bm k},t)=
{i\Delta_{\rm even} \over 2{\cal E}_{\rm even}({\bm k})}
e^{i{\cal E}_{\rm even}({\bm k})t}.
\label{eq.B9}
\end{equation}
Then, the substitution of Eq.~(\ref{eq.B9}) into Eq.~(\ref{eq.x10}) gives the second line in Eq.~(\ref{eq.y2}).
\par
\par
\section{Odd-frequency pair wavefunction in the strong-coupling limit}
\par
We derive the relation between the pair wavefunctions in the odd- and the even-frequency pairing states in the strong-coupling regime $(k_{\rm F}a_s)^{-1}\gg 1 $. In this regime, because of $\mu<0$ and $|\mu|\gg\Lambda$, the basis function in Eq.~(\ref{eq.6}) can be approximated to 
\begin{equation}
\gamma({\bm k},i\omega_n)\simeq{\rm sgn}(\omega_n).
\label{eq.B0}
\end{equation}
Then, the (1,2)-component of the thermal Green's function in Eq.~(\ref{eq.21}) is reduced to
\begin{equation}
G_{\rm odd}^{(1,2)}({\bm k},i\omega_n)
={\rm sgn}(\omega_n)
{\Delta \over \omega_n^2+{\cal E}_{\rm odd}({\bm k},\Lambda=0)^2},
\label{eq.B1}
\end{equation}
where ${\cal E}_{\rm odd}({\bm k},\Lambda)$ is given in Eq.~(\ref{eq.26}). Substituting this into Eq.~(\ref{eq.x2}), we obtain
\begin{align}
{\cal G}_{\rm odd}^{<,(1,2)}({\bm k},\omega)
&=
f(\omega)
\left[
{|\Delta| \over (\omega+i\delta)^2-{\cal E}_{\rm odd}({\bm k},\Lambda=0)^2}
\right.
\nonumber
\\
&+
\left.
{|\Delta| \over (\omega-i\delta)^2-{\cal E}_{\rm odd}({\bm k},\Lambda=0)^2}
\right].
\label{eq.B2}
\end{align}
Thus, the lesser Green's function in the time domain is given by
\begin{eqnarray}
&&{\cal G}_{\rm odd}^{<,(1,2)}({\bm k},t)
\nonumber
\\
&&=\int_{-\infty }^\infty{d\omega \over 2\pi}e^{-i\omega t}
f(\omega)
\left[
{|\Delta| \over (\omega+i\delta)^2-{\cal E}_{\rm odd}({\bm k},\Lambda=0)^2}
\right.
\nonumber
\\
&&+
\left.
{|\Delta| \over (\omega-i\delta)^2-{\cal E}_{\rm odd}({\bm k},\Lambda=0)^2}
\right].
\label{eq.B3}
\end{eqnarray}
\par
At $t=0$, Eq.~(\ref{eq.B3}) is evaluated as, by changing the variable $\omega$ as $-\omega$ in the last term,
\begin{eqnarray}
&&{\cal G}_{\rm odd}^{<,(1,2)}({\bm k},t=0)
\nonumber
\\
&&=
\int_{-\infty }^\infty{d\omega \over 2\pi}
{|\Delta| \over (\omega+i\delta)^2-{\cal E}_{\rm odd}({\bm k},\Lambda=0)^2}=0.
\label{eq.B4}
\end{eqnarray}
In obtaining the last expression, we have closed the integral path in the upper-half complex plane, and have used the analytic property of the retarded Green's function in the upper-half plane.
\par
\begin{figure}[t]
\centering
\includegraphics[width=0.45\textwidth]{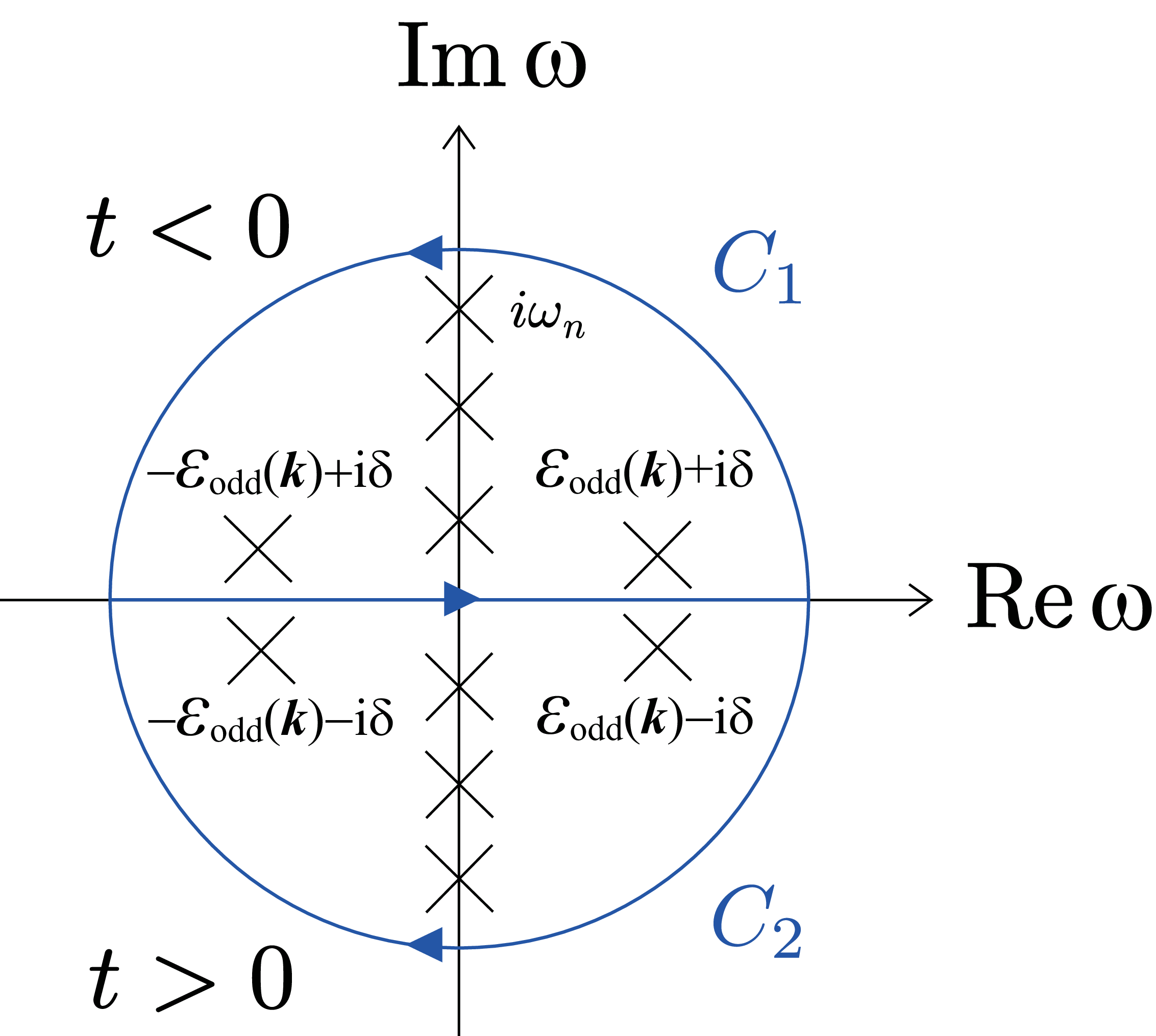}
\caption{Complex path to evaluate the $\omega$ integral in Eq.~(\ref{eq.B3}). $C_1$ ($C_2$) is chosen when $t<0$ ($t>0$).}
\label{fig11}
\end{figure}
\par
When $t\ne 0$, one can perform the $\omega$ integration in Eq.~(\ref{eq.B3}), by choosing the closed path $C_1$ ($C_2$) shown in Fig.~\ref{fig11} when $t<0$ ($t>0$). Evaluating residues at $\omega=\pm {\cal E}_{\rm odd}({\bm k},\Lambda=0) \pm i\delta$ and $i\omega_n$, we obtain
\begin{align}
&{\cal G}_{\rm odd}^{<,(1,2)}({\bm k},t\ne 0)
=
{\rm sgn}(t){i|\Delta| \over 2{\cal E}_{\rm odd}({\bm k},\Lambda=0)}
\nonumber
\\
&\times
\left[
f(-{\cal E}_{\rm odd}({\bm k},\Lambda=0)) e^{i{\cal E}_{\rm odd}({\bm k},\Lambda=0)t}
\right.
\nonumber
\\
&-
\left.
f({\cal E}_{\rm odd}({\bm k},\Lambda=0)) e^{-i{\cal E}_{\rm odd}({\bm k},\Lambda=0)t}
\right]
\nonumber
\\
&-
{\rm sgn}(t)
{2i|\Delta| \over \beta}
\sum_{n=0}^\infty
{1 \over \omega_n^2+{\cal E}_{\rm odd}({\bm k},\Lambda=0)^2} e^{-|\omega_n t|}
\nonumber
\\
&=
{\rm sgn}(t){i|\Delta| \over 2{\cal E}_{\rm odd}({\bm k},\Lambda=0)}
e^{i{\cal E}_{\rm odd}({\bm k},\Lambda=0)t}
\nonumber
\\
&+
{\rm sgn}(t)
{i|\Delta| \over \pi{\cal E}_{\rm odd}({\bm k},\Lambda=0)}I(t).
\label{eq.B5}
\end{align}
Here, we have set $T=0$ in the last expression, and
\begin{align}
I(t)
&=
-\int_0^\infty d\omega
{{\cal E}_{\rm odd}({\bm k},\Lambda=0) \over \omega^2+{\cal E}_{\rm odd}({\bm k},\Lambda=0)^2} e^{-\omega|t|}
\nonumber
\\
&=
{\rm si}({\cal E}_{\rm odd}({\bm k},\Lambda=0)|t|)
\cos({\cal E}_{\rm odd}({\bm k},\Lambda=0)|t|)
\nonumber
\\
&-
{\rm ci}({\cal E}_{\rm odd}({\bm k},\Lambda=0)|t|)
\sin({\cal E}_{\rm odd}({\bm k},\Lambda=0)|t|),
\label{eq.B7a}
\end{align}
where
\begin{equation}
{\rm si}(x)=-\int_x^\infty dy{\sin(y) \over y},
\label{eq.B7}
\end{equation}
\begin{equation}
{\rm ci}(x)=-\int_x^\infty dy{\cos(y) \over y},
\label{eq.B8}
\end{equation}
are the sine integral and cosine integral, respectively.
\par
In the strong coupling regime, since $|\Delta|\gg\Lambda$ and $|\mu|\gg\Lambda$, we can safely ignore $\Lambda$ in the gap equation~(\ref{eq.29}), as well as in the number equation~(\ref{eq.30}). The resulting expressions have the same forms as the corresponding equations~(\ref{eq.x4b}) and~(\ref{eq.x4c}) in the even-frequency case, which immediately concludes $|\Delta|=\Delta_{\rm even}$ and $\mu=\mu_{\rm even}$. Then, Eq.~(\ref{eq.B5}) can be rewritten as, by using Eq.~(\ref{eq.B9}),
\begin{align}
&{\cal G}_{\rm odd}^{<,(1,2)}({\bm k},t\ne 0)
\nonumber
\\
&=
{\rm sgn}(t)
\left[
{\cal G}_{\rm even}^{<,(1,2)}({\bm k},t\ne 0)
+
{i|\Delta|I(t) \over \pi{\cal E}_{\rm odd}({\bm k},\Lambda=0)}
\right].
\label{eq.B10}
\end{align}
When the pairing interaction is extremely strong, because $|\mu|$ and $|\Delta|$ eventually diverges, one may take ${\cal E}_{\rm odd}({\bm k},\Lambda=0)|t|\gg 1$ in Eq.~(\ref{eq.B7a}), when $t\ne 0$. Then, using the asymptotic formula,
\begin{equation}
{\rm si}(x)\cos(x)-{\rm ci}(x)\sin(x)\simeq-{1 \over x}~~~(x\gg 1),
\label{eq.B11}
\end{equation}
one finds,
\begin{equation}
I(t\ne 0)\simeq{1 \over {\cal E}_{\rm odd}({\bm k},\Lambda=0)|t|}\to 0,
\label{eq.B12}
\end{equation}
which leads to
\begin{equation}
{\cal G}_{\rm odd}^{<,(1,2)}({\bm k},t\ne 0)=
{\rm sgn}(t)
{\cal G}_{\rm even}^{<,(1,2)}({\bm k},t\ne 0).
\label{eq.B13}
\end{equation}
Thus, the pair wavefunctions $\varphi_{\rm odd}({\bm r},t)$ in Eq.~(\ref{eq.x1}) and $\varphi_{\rm even}({\bm r},t)$ in Eq.~(\ref{eq.y2}), which is related to ${\cal G}_{\rm even}^{<,(1,2)}({\bm k},t\ne 0)$ as Eq.~(\ref{eq.x10}), in the strong-coupling limit are related to each other as
\begin{equation}
\varphi_{\rm odd}({\bm r},t)=-{\rm sgn}(t)
\varphi_{\rm even}({\bm r},t).
\label{eq.B14}
\end{equation}
\par
We find from Eq.~(\ref{eq.B12}) that $|\varphi_{\rm odd}({\bm r},t)|$ becomes close to $|\varphi_{\rm even}({\bm r},t)|$, when $t\gg 1/{\cal E}_{\rm odd}({\bm k},\Lambda=0)$. Since ${\cal E}_{\rm odd}({\bm k},\Lambda=0)\ge \sqrt{|\Delta|^2+\mu^2}$ in the strong-coupling regime [where $\mu<0$], this condition may be written as
\begin{equation}
|t|\gg{1 \over \sqrt{|\Delta|^2+\mu^2}}.
\label{eq.B15}
\end{equation}
\par
\par


\begin{thebibliography}{99}
\bibitem{alt} We briefly note that the required positive $A({\bm k},\omega)$ is also obtained in the case of the model interaction where $\xi_{\bm k}^2$ in the denominator of the basis function in Eq.~(\ref{eq.6}) is replaced by $\alpha\xi_{\bm k}$ [where $\alpha\ge 1$]. However, this paper only deals with the simplest case with $\alpha=1$.
\bibitem{note2} Because of the antisymmetric property of the Fermi statistics, the superfluid order parameter must change its sign, when one exchanges all the time, spatial coordinates, as well as spin, of two fermions involved in the order parameter. Thus, since the order parameter in the even frequency $s$-wave pairing state must be an odd function with respect to the spin exchange, this pairing state is forbidden in a single-component Fermi gas. In the odd-frequency case, on the other hand, because of the antisymmetry with respect to the time variable, $s$-wave pairing symmetry is allowed in a single-component Fermi gas.
\bibitem{note22} We note that the number equation can also be obtained from the thermodynamic identity, $N=-\partial \Omega_{\rm MF}/\partial \mu$. However, the resulting number equation has an addition term ($\equiv\delta N$) to Eq.~(\ref{eq.30}), which comes from the $\mu$-derivative of $\gamma({\bm k},i\omega_n)$ involved in Eq.~(\ref{eq.24}). Since the $\mu$-dependence of the pairing interaction $V_{k,k'}$ in Eq.~(\ref{eq.4}) implicitly means that this effective interaction is produced by a more fundamental Fermi-Fermi interaction. In this sense, $\delta N$ may be interpreted as a correction to the mean-field number equation~(\ref{eq.30}) by this interaction. Thus, we deal with the mean-field type number equation~(\ref{eq.30}), for simplicity.
\bibitem{note3} A dip structure is seen in Fig.~\ref{fig9}(c) slightly above the unitary limit, when $\Lambda/\epsilon_{\rm F}=0.1 \ll 1$. Regarding this, we briefly note that it comes from the fact that the function $f\left(\sqrt{(\varepsilon_{\bm k}-\mu)^2+\Lambda^2} + i\nu_n/2\right)$ in Eq.~(\ref{eq.32}) is reduced to the Bose distribution function $n_{\rm B}\left(\sqrt{(\varepsilon_{\bm k}-\mu)^2+\Lambda^2}\right)$ when $n$ in $\nu_n=2\pi nT$ is an odd number. Then $n_{\rm B}\left(\sqrt{(\varepsilon_{\bm k}-\mu)^2+\Lambda^2}\right)$ enhances pairing fluctuations described by $\Pi(q)$ in Eq.~(\ref{eq.32}) when $\mu\simeq 0$ and $\Lambda/\epsilon_{\rm F}\ll 1$, leading to this dip structure. However, since $f\left(\sqrt{(\varepsilon_{\bm k}-\mu)^2+\Lambda^2} + i\nu_n/2\right)$ relies on the details of the basis function $\gamma({\bm k},i\omega_n)$ in Eq.~(\ref{eq.6}), this fine structure would be peculiar to the present model, we are considering in this paper.
\bibitem{note4} Equation~(\ref{eq.46}) is slightly different from Eq.~(\ref{eq.40}). (That is, $\varepsilon_{\rm F}$ in the latter equation is replaced by $\mu$ in the former.) This is simply because, while the number equation~(\ref{eq.30}) in the BCS-Eagles-Leggett theory always gives $\mu=\varepsilon_{\rm F}$ when $\Delta=0$, the NSR number equation~(\ref{eq.34}) involving effects of pairing fluctuations gives $\mu<\varepsilon_{\rm F}$ even at $T=0$. As a result, the threshold interaction strength seen in Fig.~(\ref{fig9}) is slightly larger than Fig.~\ref{fig4}.
\end{thebibliography}
\end{document}